\RequirePackage{fix-cm}
\documentclass[smallextended]{svjour3}       
\smartqed  
\usepackage{graphicx}
\usepackage{amsmath, bm}
\usepackage{amsfonts}
\usepackage{booktabs}
\usepackage{tabularx}
\usepackage[table]{xcolor}  
\usepackage{multirow}       
\usepackage{array}          
\usepackage{subfigure}
\usepackage{algorithm}
\usepackage[noend]{algpseudocode}
\usepackage{xcolor}
\usepackage{todonotes}
\usepackage{siunitx}
\usepackage{verbatim}
\usepackage{multirow} 
\usepackage{hyperref}
\usepackage{tikz}
\usetikzlibrary{shapes.geometric, arrows.meta, positioning,patterns,calc}
\hypersetup{hidelinks,
	colorlinks=true,
	allcolors=black,
	pdfstartview=Fit,
	breaklinks=true}

\newcommand{\dan}[1]{\textcolor{red}{{\textbf{Dan:}} #1}}
\newcommand{\jinlong}[1]{\todo[color=red!30, inline]{\textbf{Jinlong:} #1}}
\newcommand{\shu}[1]{\todo[color=green!30, inline]{\textbf{Shu:} #1}}
\newcommand{\huzaifa}[1]{\todo[color=yellow!30, inline]{\textbf{Huzaifa:} #1}}
\newcommand{\jingquan}[1]{\todo[color=purple!30, inline]{\textbf{Jingquan:} #1}}
\newcommand{\updatedText}[1]{\textcolor{black}{#1}}
\newcommand{\updatedTextRebutalTwo}[1]{\textcolor{black}{#1}}
\definecolor{hanblue}{rgb}{0.27, 0.42, 0.81}
\begin{document}

\clearpage
\newpage

\title{MBD-NODE: Physics-informed data-driven modeling and simulation of constrained multibody systems}


\author{Jingquan Wang        \and
        Shu Wang \and 
         Huzaifa Mustafa Unjhawala \and
         Jinlong Wu \and
         Dan Negrut
}

\authorrunning{Wang, Wang, Unjhawala, Wu, Negrut} 

\institute{Jingquan Wang \at
Department of Mechanical Engineering, University of Wisconsin-Madison, 1513 University Avenue, 53706, Madison, USA\\
              \email{jwang2373@wisc.edu}           
           \and
           Shu Wang \at
           Department of Mechanical Engineering, University of Wisconsin-Madison, 1513 University Avenue, 53706, Madison, USA\\
           \email{swang579@wisc.edu}
           \and 
           Huzaifa Mustafa Unjhawala \at
    Department of Mechanical Engineering, University of Wisconsin-Madison, 1513 University Avenue, 53706, Madison, USA\\
    \email{unjhawala@wisc.edu}
    \and 
    Jinlong Wu \at
    Department of Mechanical Engineering, University of Wisconsin-Madison, 1513 University Avenue, 53706, Madison, USA\\
    \email{jinlong.wu@wisc.edu}
    \and
    Dan Negrut \at
    Department of Mechanical Engineering, University of Wisconsin-Madison, 1513 University Avenue, 53706, Madison, USA\\
    \email{negrut@wisc.edu}
}

\date{Received: date / Accepted: date}

\maketitle

\begin{abstract}

We describe a framework that can integrate prior physical information, e.g., the presence of kinematic
constraints, to support data-driven simulation in multi-body dynamics. Unlike other approaches, e.g., Fully-connected Neural Network (FCNN) or Recurrent Neural Network (RNN)-based methods that are used to model the system states directly, the proposed approach embraces a Neural Ordinary Differential Equation (NODE)
paradigm that models the derivatives of the system states. A central part of the proposed methodology is its
capacity to learn the multibody system dynamics from prior physical knowledge and constraints combined with data inputs. This learning process is facilitated by a constrained optimization approach, which
ensures that physical laws and system constraints are accounted for in the simulation process. The models, data, and code for this work are publicly available as open source at \href{https://github.com/uwsbel/sbel-reproducibility/tree/master/2024/MNODE-code}{https://github.com/uwsbel/sbel-reproducibility/tree/master/2024/MNODE-code}.

\keywords{Multibody dynamics \and Neural ODE \and Constrained dynamics \and Scientific machine learning}

\subclass{34C60
 \and 37M05}
\end{abstract}

\section{Introduction}
\label{intro}

This contribution is concerned with using a data-driven approach to characterize the dynamics of multibody systems. Recently, data-driven modeling methods have been developed to characterize multibody dynamics systems based on neural networks. For instance, fully connected neural networks (FCNN) offer a straightforward way to model multibody dynamics, mapping input parameters (potentially including time for non-autonomous systems), to state values \cite{Choi_DDSdd57a6f7c2064450bc1f79231ef67414,pan2021vehicle_mbd}. FCNNs act as regressors or interpolators, predicting system states for any given time and input parameter. The method's simplicity allows rapid state estimation within the training set range. However, this approach requires expanding the input dimension to accommodate, for instance, changes in initial conditions, leading to an exponential increase in training data. An alternative method combines fixed-time increment techniques with principal component analysis (PCA) \cite{efficient_PCA} to reduce training costs and data requirements. This method outputs time instances at fixed steps and employs PCA for dimensionality reduction. However, this results in discontinuous dynamics, limiting output to discrete time steps. Another approach uses dual FCNNs, one modeling dynamics, and the other estimating errors, to enhance accuracy while reducing computational costs \cite{HAN_DNN2021113480}. A limitation of FCNN is the strong assumption of independent and identically distributed \updatedText{(i.i.d.) data \cite{FCNN_iid}} required by the approach, which limits its ability to generalize, particularly in terms of extrapolation across time and phase space.

Time series-based approaches (e.g., LSTM~\cite{hochreiter1997lstm}) have also been investigated for the data-driven modeling of multibody dynamics systems. These methods learn to map historical state values to future states, but they are inherently discrete and tied to specific time steps. For example, RNNs have been directly employed for predicting subsequent states for a drivetrain system~\cite{RNNMBD} and a railway system \cite{2019RNN_railway}. The time-series approach works well for systems with strong periodic patterns, a scenario in which the approach yields high accuracy. RNN turned out to be challenged by systems with more parameters than just time (e.g., accounting of different initial conditions). To mitigate its shortcomings, a combination of 3D Convolutional Neural Network(3DCNN), FCNN, and RNN has been used for tracked-vehicle system behavior prediction \cite{YE_MBSNET2021107716}.  More examples using FCNN and RNN can be found in the review paper \cite{hashemi2023multibody_review}.

Neural ordinary differential equations (NODE), recently introduced in \cite{Chen_NODE_2018}, provide a framework that offers a more flexible data-driven approach to model continuous-time dynamics. Unlike traditional regression-based models which assume a discretized time step (which is often fixed a priori), NODE employs a continuous way of modeling an ordinary differential equation (ODE) and allows flexible discretization in the numerical simulation. This makes NODE capable of dealing with data in different temporal resolutions and emulating physical systems in continuous time. In terms of applications, NODE has seen good success in engineering applications, such as chemical reactions \cite{OWO_CHEMNODE_2022,chem_node2}, turbulence modeling \cite{portwood2019turbulence}, spintronic dynamics, \cite{chen2022forecasting}, and vehicle dynamics problems \cite{quaglino2019snode}.
\updatedText{Recent work has significantly advanced our understanding of NODE-based approaches, providing analyses of convergence \cite{node_convergence}, robustness \cite{node_robust}, and generalization ability \cite{node_general1,node_general2,node_general3}}.



Conservation laws are important to be obeyed by the model used to characterize the response of the system. Hamiltonian Neural Networks (HNN)\cite{HNN}, inspired by Hamiltonian mechanics, can factor in exact conservation laws by taking generalized positions and momenta as inputs to model the system's Hamiltonian function. However, a HNN requires data in generalized coordinates with momentum, posing practical challenges, especially in multibody dynamics (MBD) systems as it necessitates the transformation of complex, often high-dimensional system dynamics into a reduced set of generalized coordinates. This transformation can be both computationally intensive and prone to inaccuracies, especially when dealing with intricate mechanical systems involving multiple interacting components whose dynamics is constrained through mechanical joints. In addition, even if a MBD system is non-dissipative, it can still represent a non-separable Hamiltonian system, for which the integration process is more complicated, often necessitating implicit methods. Various follow-up works have expanded upon HNN, addressing systems with dissipation \cite{zhong2019dissipativeHN,DissipativeHNN}, generative networks \cite{toth2019hamiltonian_gen}, graph neural networks \cite{sanchez2019hamiltonian_graph}, symplectic integration \cite{DAVID_symplectic_HNN,chen2020symplecticRNN,HNN_IET}, non-separable Hamiltonian system \cite{ssinn2020_symplectic_HNN}, and the combination with probabilistic models \cite{bacsa2023symplectic_nature}.

To address the limitations of HNN, Lagrangian Neural Networks (LNN)~\cite{LNN,lutter2019deeplnn}, have been proposed to leverage Lagrangian mechanics. LNN models the system Lagrangian, with second-order state derivatives derived from the Euler-Lagrange equation. This approach also conserves the total energy and applies to a broader range of problems. However, it is computationally intensive and sometimes ill-posed due to its reliance on the inverse Hessian. Subsequent works on LNN have explored various aspects, such as including constraints \cite{finzi2020simplifying_HNNLNN_CON}, extended use with graph neural networks \cite{bhattoo2023lnn_graph}, and model-based learning \cite{lnn_hnn_contact,Lagrangian_model_based}.

In this study, our primary objective is to learn the dynamics of multibody systems from system states data using a \updatedText{NODE}-based approach. We also explore the process of incorporating prior physical information, such as kinematic constraints, into the numerical solution through the use of a constrained optimization method, in conjunction with standard \updatedText{NODEs}. We compare the performance of the proposed approach with existing methodologies on various examples. Our contributions are as follows:
\begin{itemize}
    \item We propose a method called Multibody NODE (\updatedTextRebutalTwo{MBD-NODE}), by applying \updatedText{NODE} to the data-driven modeling of general MBD problems, and establishing a methodology to incorporate known physics and constraints in the model.
    \item We provide a comprehensive comparison of the performance of \updatedTextRebutalTwo{MBD-NODE} with several other methods that have been applied to MBD problems.
    \item We build a series of MBD test problems; providing an open-source code base consisting of several data-driven modeling methods (i.e., FCNN, LSTM, HNN, LNN, \updatedTextRebutalTwo{MBD-NODE}); and curating a well-documented summary of their performances.
\end{itemize}

\section{Methodology}
\label{sec:methodology}

\subsection{Multibody System Dynamics}
MBD is used in many mechanical engineering applications to analyze systems composed of interconnected bodies. \updatedText{Here we rely on the general form of the MBD problem \cite{MBD}, which accounts for the presence of constraint equations using Lagrange multipliers in the equations of motion}:
\begin{equation}
\label{eqn:MBD}
\begin{bmatrix}
\mathbf{M} & \mathbf{\Phi}^T_\mathbf{q} \\
\mathbf{\Phi_q} & 0 \\
\end{bmatrix}
\begin{bmatrix}
\ddot{\mathbf{q}} \\
\lambda \\
\end{bmatrix}
=
\begin{bmatrix}
\mathbf{F_e} \\
\mathbf{\gamma_c}\\
\end{bmatrix},
\end{equation}
where $\mathbf{M}$ represents the mass matrix; $\mathbf{\Phi_\mathbf{q}}$ is the constraint Jacobian matrix; $\mathbf{q}$ denotes the vector of system states (generalized coordinates); $\ddot{\mathbf{q}}$ denotes the acceleration vector of the system; $\lambda$ represents the Lagrangian multipliers; $\mathbf{F_e}$ is the combined vector of generalized external forces and quadratic velocity terms; \updatedText{and $\mathbf{\gamma_c}$ is the right hand side of the kinematic constraint equations at the acceleration level}. In practice, the set of differential-algebraic equations (DAE) in Eq.~\eqref{eqn:MBD} can be numerically solved by several methods, see, for instance, \cite{bauchau2008review}.

\subsection{Neural Ordinary Differential Equations for Multibody System Dynamics}
\updatedText{\subsubsection{Neural Ordinary Differential Equation (NODE)}}
NODE represents a class of deep learning models that train neural networks to approximate the unknown vector fields in ordinary differential equations (ODEs) to characterize the continuous-time evolution of system states. Given a hidden state $\mathbf{z}(t), \mathbf{z}\in \mathbb{R}^{n_z}$ at time $t$, the NODE is defined by the following equation:
\begin{equation}\label{eqn: NODE}
    \frac{d\mathbf{z}(t)}{dt}=f(\mathbf{z}(t),t;\ \mathbf{\Theta}),
\end{equation}
where $f:\mathbb{R}^{n_z}\times \mathbb{R}^{+}\to\mathbb{R}^{n_z}$ corresponds to a neural network parameterized by ${\mathbf{\Theta}}$. For an arbitrary time $t>0$, the state $\mathbf{z}(t)$ can be obtained by solving an initial value problem (IVP) through the forward integration:
\begin{equation}
    \mathbf{z}(t) = \mathbf{z}(0)+\int_0^t f(\mathbf{z}(\tau),\tau;\ \mathbf{\Theta})d\tau = \Phi(\mathbf{z}(0),f, t),
\label{eqn: NODE_IVP}
\end{equation}
where $\Phi$ denotes an ODE solver.

\updatedText{NODE} \cite{Chen_NODE_2018,adap_adjoint_NODE} provides an efficient approach of calibrating the unknown parameters $\mathbf{\Theta}$ based on some observation data $\mathbf{z}(t_i)$ for $i=1,2,...,n$. Note that the $t_i$ time \updatedText{steps} do not have to be equidistant and thus one has flexibility in choosing numerical integrators for the forward integration in Eq. \eqref{eqn: NODE_IVP}.



\updatedText{\subsubsection{Extensions of Neural Ordinary Differential Equation}}
In the modeling of dynamical systems, it is quite common for equations to include parameters that significantly influence the system's behavior, such as the Reynolds number in the Navier-Stokes equations or design parameters in MBD, e.g., lengths, masses, material properties. Enhancing NODEs to accommodate such variations would enable the simultaneous learning of a wide range of dynamics. A practical way to achieve this augmentation is to incorporate these parameters directly into the neural network's inputs, known as PNODE that is suggested in \cite{PNODE}:


\begin{equation}
\frac{d\mathbf{z}(t)}{dt} = f(\mathbf{z}(t), t,\bm{\mu};\ \mathbf{\Theta}),
\end{equation}
where \(\bm{\mu} = (\mu_1, \mu_2, \dots, \mu_{n_{\bm{\mu}}})^T \in \mathbb{R}^{n_{\bm{\mu}}}\) is the parameter vector that can help better characterize the system, \(f: \mathbb{R}^{n_z }\times \mathbb{R}^{+}\times \mathbb{R}^{n_{\bm{\mu}}} \to \mathbb{R}^{n_z}\) is the neural network. 

For the systems whose governing equations are second-order, we can use the second-order neural ordinary differential equation (SONODE)~\cite{norcliffe2020secondNODE,SONODE_optimizer} to model them. Given an augmented state $\mathbf{Z}(t)=(\mathbf{z}(t), \mathbf{\dot z}(t))^T$, the SONODE is defined as:
\begin{equation}\label{eqn: Second_NODE}
\frac{d\mathbf{Z}(t)}{dt} = f(\mathbf{Z}(t), t;\ {\mathbf{\Theta}}),
\end{equation}
where $f: \mathbb{R}^{2n_z}\times \mathbb{R}^{+} \to\mathbb{R}^{2n_z}$ is a neural network parameterized by $\mathbf{\Theta}$. 

{\subsubsection{\updatedText{Multibody Dynamics NODE (\updatedTextRebutalTwo{MBD-NODE})}}
Based on the above PNODE and SONODE, we extend the approach to make the NODE work with external inputs \updatedText{like external generalized forces, thus better fitting the MBD framework. Given the set of generalized coordinates $\mathbf{Z}(t,\bm{\mu})=(\mathbf{z}^T(t,\bm{\mu}), \mathbf{\dot z}^T(t,\bm{\mu}))^T \in \mathbb{R}^{2n_z}$, the \updatedTextRebutalTwo{MBD-NODE} is defined as}

\updatedText{
\begin{equation}\label{eqn: MNODE}
\frac{d\mathbf{Z}(t,\bm{\mu})}{dt} = f(\mathbf{Z}(t,\bm{\mu}),\mathbf{u}(t), t,\bm{\mu};\ \mathbf{\Theta}),
\end{equation}}
where: 
\updatedText{
\begin{equation}
    \mathbf{Z}(0,\bm{\mu}) = (\mathbf{z}^T(0,\bm{\mu}), \mathbf{\dot z}^T (0,\bm{\mu}))^T,
\end{equation}
}
\updatedText{are the initial values} for the MBD;

\updatedText{
\begin{equation}
    \mathbf{z}(t,\bm{\mu}) = (z^1(t,\bm{\mu}),...,z^{n_z}(t,\bm{\mu}))^T\in \mathbb{R}^{n_z},  
\end{equation}}
\updatedText{are the generalized positions};
\updatedText{
\begin{equation}
    \mathbf{\dot z}(t,\bm{\mu}) = (\dot z^1(t,\bm{\mu}),\dots,\dot z^{n_z}(t,\bm{\mu}))^T\in \mathbb{R}^{n_z},  
\end{equation}}
\updatedText{are the generalized velocities};
\updatedText{
\begin{equation}
    \mathbf{u}(t)=(u^1(t),...,u^{n_u}(t))^T\in \mathbb{R}^{n_u},  
\end{equation}}
\updatedText{are the $n_u$ external loads} like force/torque applied to the MBD at time $t$ (note that time $t$ can be included in the input $\mathbf{u}(t)$);
\updatedText{
\begin{equation}
    \bm{\mu} = (\mu_1, \mu_2, \dots, \mu_{n_{\bm{\mu}}})^T \in \mathbb{R}^{n_{\bm{\mu}}},  
\end{equation}}
are problem-specific parameters, and 
\updatedText{
\begin{equation}
    f: \mathbb{R}^{2n_z}\times \mathbb{R}^{n_u}\times\mathbb{R}^{n_{\bm \mu}}\to \mathbb{R}^{2n_z},  
\end{equation}}
 is the neural network parameterized by $\mathbf{\Theta}$ with $2n_z+n_u+n_{\mu}$ dimensional input.
For the forward pass to solve the initial value problem for $\mathbf{Z}(t,\bm \mu)$, we can still use the integrator $\Phi$ that:

\begin{equation}
    \mathbf{Z}(t,\bm \mu)=\mathbf{Z}(0,\bm \mu)+\int_0^t f(\mathbf{Z}(\tau,\mu),\mathbf{u}(\tau),t,\bm{\mu};\ \mathbf{\Theta})d\tau=\Phi(\mathbf{Z}(0,\bm \mu),f,\updatedText{\mathbf{u}},t).
\end{equation}

For the backpass of the \updatedTextRebutalTwo{MBD-NODE}, we can use the backpropagation or the adjoint method to design the corresponding adjoint state based on the property of second-order ODE \cite{norcliffe2020secondNODE,SONODE_optimizer}. We finally choose to use backpropagation, a step analyzed in detail in the next section~\ref{sec: loss}, which touches on the construction of loss function and optimization.

\begin{figure}[h]
    \centering
    \includegraphics[width=6cm]{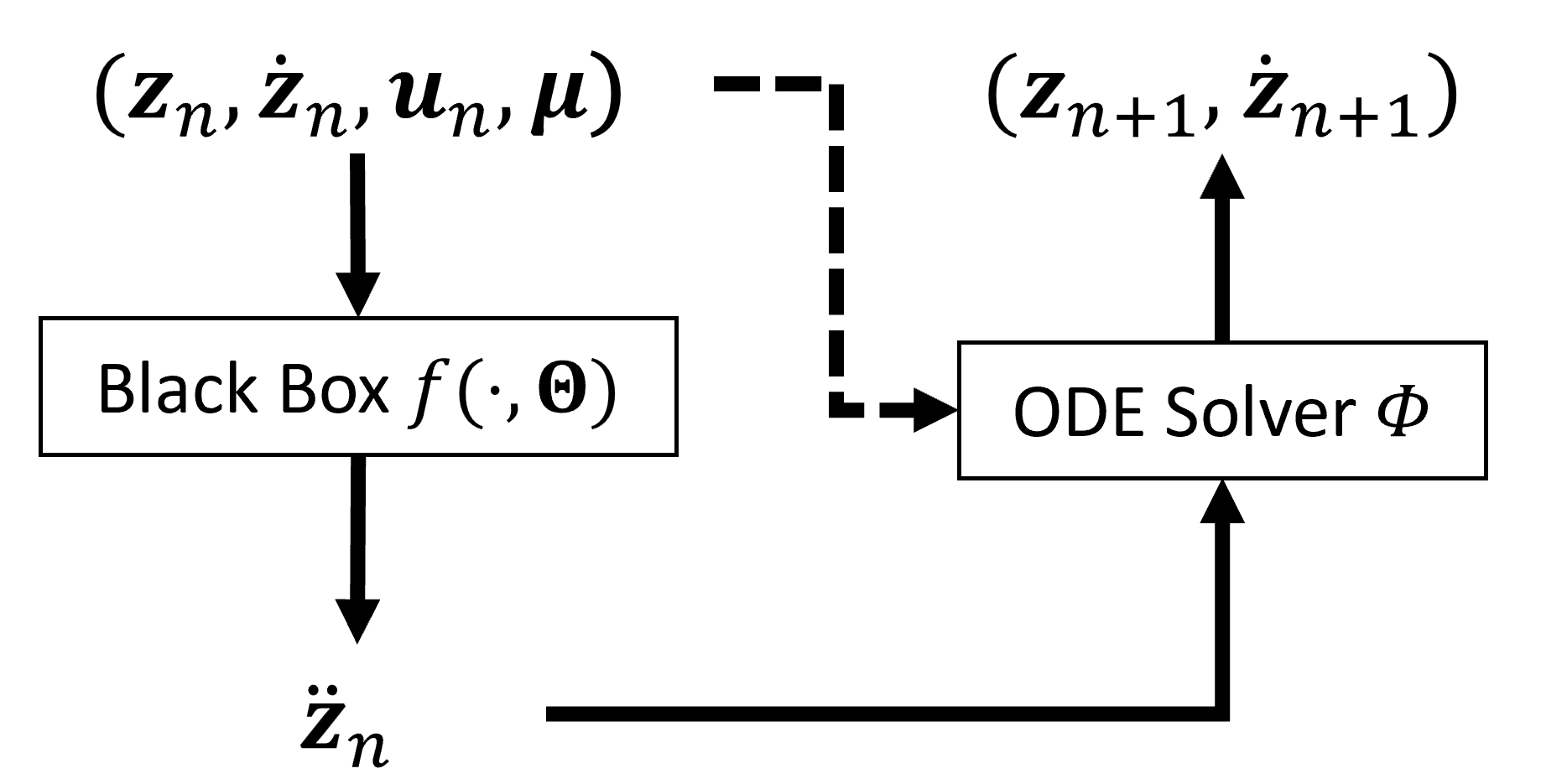}
    \caption{The discretized forward pass for \updatedTextRebutalTwo{MBD-NODE} for general MBD.}
    \label{fig: MNODE_con}
\end{figure}

\begin{figure}[h]
    \centering
    \includegraphics[width=6cm]{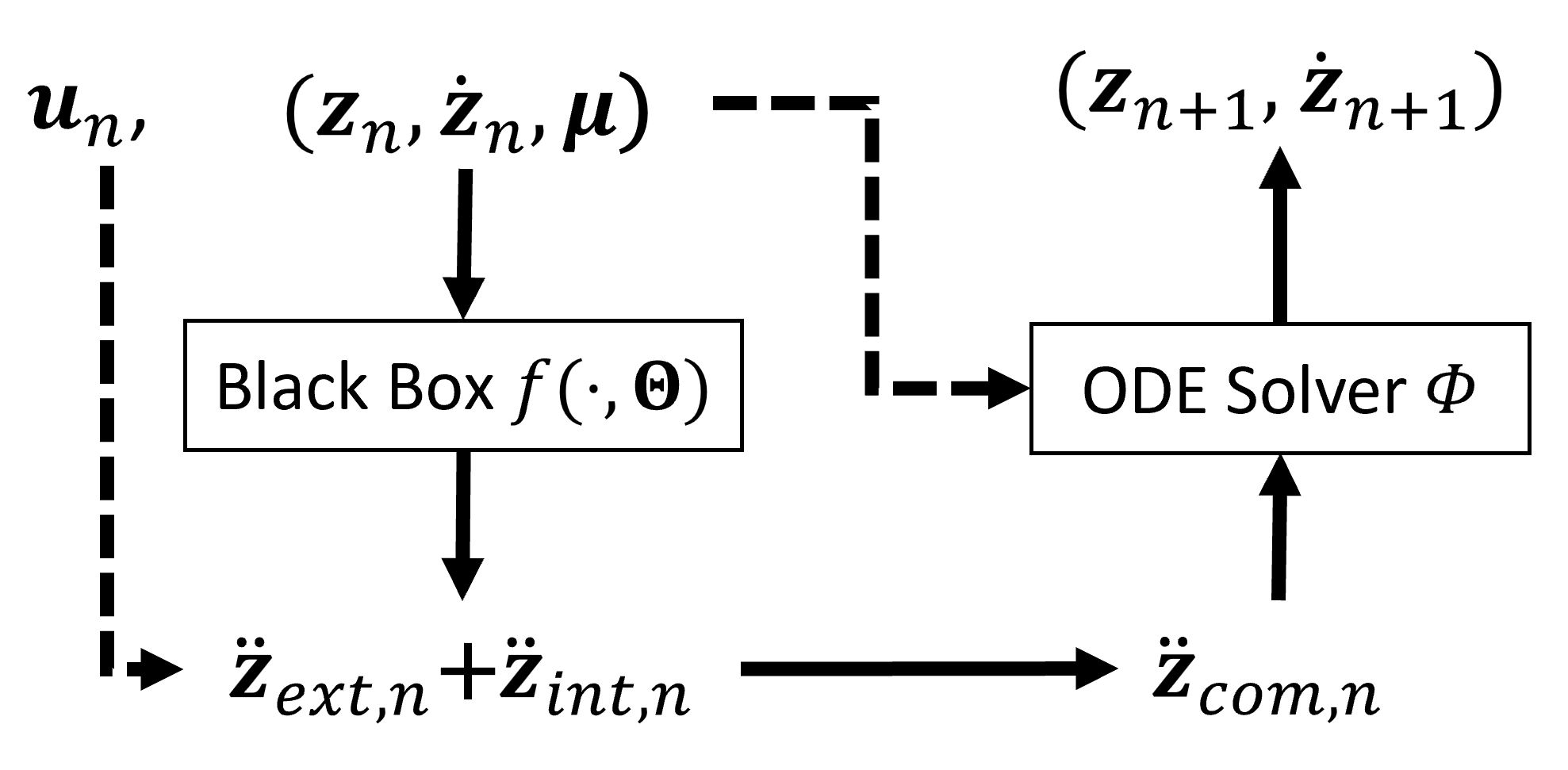}
    \caption{The discretized forward pass for \updatedTextRebutalTwo{MBD-NODE} for general MBD without hard \updatedText{constraints} which means the external force/torque could be directly added to the acceleration from the NODE that models the internal acceleration without additional input channel for external force. Here $\mathbf{\ddot z}_{ext,n}$ is the acceleration caused by the external input and $\mathbf{\ddot z}_{int,n}$ is the internal acceleration predicted by \updatedTextRebutalTwo{MBD-NODE}.}
    \label{fig: MNODE_un}
\end{figure}
Figures \ref{fig: MNODE_con} and \ref{fig: MNODE_un} show the discretized version of the forward pass for the MBD \updatedText{ with and without constraints. The constraint-related formulations are discussed in the Section \ref{sec: constraints}.} Within the \updatedTextRebutalTwo{MBD-NODE} framework, the initial state of the system is processed using an ODE solver, evolving over a time span under the guidance of a neural network's parameters. This neural network is trained to determine the optimal parameters that best describe the system's dynamics. This continuous approach, in contrast to discrete-time models, often results in enhanced flexibility, efficiency, and good generalization accuracy. Based on the notation used for the definition of \updatedTextRebutalTwo{MBD-NODE} in equation (\ref{eqn: MNODE}), we employ the corresponding three-layer neural network architecture in \updatedText{Table \ref{tab:model_architecture}; the activation function can be Tanh and ReLU \cite{ActivationFC}, and the initialization strategy used is that of Xavier \cite{xavier_ini}, and Kaiming \cite{kaiming_ini}.}

\begin{table}[htbp]
\centering
\caption{\updatedTextRebutalTwo{MBD-NODE} Architecture}
\label{tab:model_architecture}
\begin{tabular}{@{}lccc@{}}
\toprule
Layer          & Number of Neurons & Activation Function & Initialization \\
\midrule
Input Layer     & $2n_z+n_{\bm \mu}+n_u$& [Tanh,ReLU]& [Xavier, Kaiming]\\
Hidden Layer 1  & $d_{\text{width}}$               & [Tanh,ReLU] & [Xavier, Kaiming] \\
Hidden Layer 2  & $d_{\text{width}}$               &  [Tanh,ReLU] & [Xavier, Kaiming]               \\
Output Layer    & \updatedText{$2n_z$}                & -  & [Xavier, Kaiming]                 \\
\bottomrule
\end{tabular}
\end{table}

\updatedText{\subsubsection{Loss Function and Optimization without Constraints}}\label{sec: loss}

First, we discuss the loss function for the MBD without constraints. Without loss of generality, we assume there are no additional parameters $\bm \mu$  for notation simplicity. For a given initial state $\mathbf{Z}_0=(\mathbf{z}_0,\mathbf{\dot z}_0)$, assume the system's next state $\mathbf{Z}_1=(\mathbf{z}_1,\mathbf{\dot z}_1)$ is obtained with the integrator $\Phi$ used over a time interval {\updatedText{$\Delta t$, which can be one or several numerical integration time steps}}. The loss function used for the \updatedTextRebutalTwo{MBD-NODE} describes the mean square error (MSE) between the ground truth state and the predicted state:
\begin{equation}\label{eqn:loss}
\text{L}(\mathbf{\Theta}) = \lVert \Phi(\mathbf{Z}_0,f,\Delta t) - \mathbf{Z}_1\rVert^2_2 =\lVert \hat{\mathbf{Z}}_1- \mathbf{Z}_1 \rVert^2_2=\lVert (\hat{\mathbf{z}}_1,\hat{\dot{\mathbf{z}}}_1)- (\mathbf{z}_1,\mathbf{\dot z}_1) \rVert^2_2 \; ,
\end{equation}
where $\hat{\mathbf{Z}}_1=({\hat{\mathbf{z}}}_1^T,{\hat{\dot{\mathbf{z}}}}_1^T)^T$ is the \updatedText{predicted state by integration with the derivatives from \updatedTextRebutalTwo{MBD-NODE}}.

For a trajectory of states $\mathbf{z}_0,\mathbf{z}_1,...,\mathbf{z}_T$, \updatedText{the common way \cite{Chen_NODE_2018} is to treat the first state as initial condition and all other states as targets,} so the loss function could be defined as that:

\updatedText{
\begin{equation}\label{eqn:loss_traj1}
\text{L}(\mathbf{\Theta}) = \sum_{i=0}^{T-1}\lVert \Phi(\mathbf{Z}_0,f,\Delta t_i) - \mathbf{Z}_{i+1}\rVert^2_2 =\sum_{i=0}^{T-1}\lVert \hat{\mathbf{Z}}_{i+1}- \mathbf{Z}_{i+1} \rVert^2_2.
\end{equation}
}
The training phase is to refine the neural network's parameters, ensuring that the predicted states mirror the true future states, which yields the optimization problem

 \begin{equation}\label{eqn: opt_unconstraints}
     \mathbf{\mathbf{\Theta}}^*=\underset{\mathbf{\Theta}}{\operatorname{argmin}}\quad\text{L}(\mathbf{\Theta}).
 \end{equation}

Similar to most deep learning models, the parameter optimization of \updatedTextRebutalTwo{MBD-NODE} can be conducted by backpropagation via stochastic gradient descent (SGD). The key for \updatedText{NODE}-based frameworks is that the objective is to fit the entire trajectory, which necessitates the storage of intermediate gradients through the integration of the whole trajectory by backpropagation. This process needs a memory cost of $O(NCL)$, where $N$ represents the number of time steps of the trajectory, $C$ is the number of neural network calls per integration step, and $L$ is the number of layers in the NODE. To solve this, adjoint methods \cite{Chen_NODE_2018} and their adaptive enhancements \cite{adap_adjoint_NODE} were implemented in the NODE-based model, achieving gradient approximation with only $O(L)$ memory costs. Further, specialized adjoint methods have been proposed for symplectic integrators \cite{adjoint_symplectic} and SONODE \cite{norcliffe2020secondNODE}, each tailored for specific applications.

In practice, optimizing parameters to fit lengthy trajectories from highly nonlinear dynamics did not work well for our problems. To address this, we partition the long trajectory, consisting of \(n\) states, into \([n/w]+1\) shorter sub-trajectories of length \(w\). The training process is then moved to these sub-trajectories. While this strategy makes the optimization easier, it may slightly impair the neural network's capacity for long-term prediction. In practice, we set $w$ to be 1 for our numerical test, and we didn't find the obvious loss of capacity for long-term prediction. 
\updatedText{In this case, the loss function will be the sum of the loss of each sub-trajectory:
\begin{equation}\label{eqn:loss_traj2}
    \text{L}(\mathbf{\Theta}) = \sum_{i=0}^{T-1}\lVert \Phi(\mathbf{Z}_i,f,\Delta t_i) - \mathbf{Z}_{i+1}\rVert^2_2 =\sum_{i=0}^{T-1}\lVert \hat{\mathbf{Z}}_{i+1}- \mathbf{Z}_{i+1} \rVert^2_2.
    \end{equation}
}

The constant $C$, the number of neural network calls per integration step, depends on the integrator used. For the Runge-Kutta \updatedText{4th} order method, the $C$ is \updatedText{4} because we need to evaluate the acceleration at intermediate states during one step of integration, while for the Forward Euler method, $C$ is 1. Also, for the implicit solvers, $C$ is the same as their explicit version because, in the training stage, we already have the next state.

Given these considerations, the memory cost for optimization via backpropagation remains within acceptable limits. For the system subject to constraints (discussed in section~ \ref{sec: constraints}), the adjoint method may not align with the \updatedText{used} method. Upon reviewing recent literature, we found no instances of the adjoint methods being applied to constrained problems. Based on these, we finally choose backpropagation to optimize the neural network. The main process for training the \updatedTextRebutalTwo{MBD-NODE} without constraints is summarized in the Algorithm \ref{alg:training_nocon} of Appendix \ref{sec:alg}.

\updatedText{\subsubsection{Loss Function and Optimization with Constraints}}\label{sec: constraints}
\updatedText{For MBD problems, accounting for constraints in the evolution of a system is imperative. These constraints capture not only physical design attributes, e.g. a spherical joint requires two points to coincide, but also factor in conservation laws, e.g., energy, numerical Hamiltonian. Accounting for these constraints is important in MBD. However, integrating constraints within deep neural network models is still an open problem}, and further research and exploration are necessary.

From a high vantage point, constraints fall in one of two categories. Holonomic constraints depend solely on the coordinates without involving the latter's time derivatives, and can be represented as \(\phi(\mathbf{z}, t) = 0\). Nonholonomic constraints, involving the time derivatives of the coordinates and cannot be time-integrated into a holonomic constraint, are denoted as \(\phi(\mathbf{z}, \mathbf{\dot{z}}, t) = 0\).

Additionally, constraints can be categorized based on their temporal dependency. Scleronomic constraints, or geometric constraints, do not explicitly depend on time and are expressed as \(\phi(\mathbf{z}) = 0\). In contrast, rheonomic constraints, which depend on time, can also be framed in the form \(\phi(\mathbf{z}, t) = 0\).

In summary, using the same notation in the above Section \ref{sec: loss}, the MBD constraints can be expressed in a generalized form \(\phi(\mathbf{z}, \mathbf{\dot{z}}, \mathbf{u},\bm{\mu}) = 0\), and the optimization problem solved can be posed as:
\begin{align}\label{eqn:hard_constraints}
\min_{\mathbf{\Theta}} \quad & \text{L}(\mathbf{\Theta}) \\
\text{s.t.}
& \phi_i(\mathbf{z}, \mathbf{\dot{z}}, \mathbf{u},\bm{\mu})=0,\forall \updatedText{ (\mathbf{z}, \mathbf{\dot{z}}, \mathbf{u},\bm{\mu})\in \mathbb{R}^{2n_z}\times\mathbb{R}^{n_u}\times\mathbb{R}^{n_{\bm \mu}}}\cap \Omega,i=1,..,n_{c},
\end{align}
where $\Omega$ is the areas from the prior physical knowledge that the MBD should have constraints.

There are two common ways to handle hard constraints. One is to relax this hard \updatedText{constrained} problem to a soft constraint problem by adding the constraints to the loss function as a penalty term \cite{fortin2000augmented,PINN_constraints,lim2022unifying,PINODE_constraints-djeumou22a}. The loss function then becomes
\begin{equation}\label{eqn: aug}
\text{J}(\mathbf{\Theta}) = \text{L}(\mathbf{\Theta}) + \sum_i g_i(\phi_i(\mathbf{z}, \mathbf{\dot{z}}, \mathbf{u},\bm \mu)) \; ,
\end{equation}
where $g_i$ represents the function for the \( i \)-th constraint, typically comprising a quadratic and a linear term, as is common in the well-known augmented Lagrangian method~\cite{fortin2000augmented,PINODE_constraints-djeumou22a,PINN_constraints}. The primary advantage of this approach is its ease of implementation, requiring only the addition of constraints as a regularization term. However, there are several drawbacks to it: the optimization process may not always converge, and the use of regularization can often diminish accuracy. Most critically, the constraints are applied exclusively within the training set's phase space, rendering them ineffective in domains beyond this phase space.

The alternative is to enforce the constraints in both the training and inference stages without adding a constraint loss term. \updatedText{Based on a coordinates partition technique \cite{Wehage82}, we denote the minimal (or independent) coordinates as $\mathbf{Z}^{M}$ and the dependent coordinates as $\mathbf{Z}^{D}$. Then, the dependent coordinates can be obtained from the independent coordinates and the prior knowledge of constraints:}
\updatedText{
\begin{equation}\label{eqn:con}
    \mathbf{Z}^{D}=\phi^{-1}(\mathbf{Z}^{M}, \bm{\mu}), 
\end{equation}}
\updatedText{where $\phi^{-1}$ is defined as the inverse function that maps the value of minimal coordinates to the dependent coordinates, and typically does not have a closed form yet it can be evaluated given $\mathbf{Z}^{M}$.}
\updatedText{If the MBD system has $n_z$ generalized coordinates and \updatedText{$n_c$} position constraints ($n_z-n_c=$ DOF), we build the \updatedTextRebutalTwo{MBD-NODE} only with the minimal coordinates $\mathbf{Z}^{M}$. Depending on whether we have the ground truth data of the dependent coordinates $\mathbf{Z}^{D}$, the training stage could be divided into two cases:}

\updatedText{(1) If we have the complete information of the dependent coordinates $\mathbf{Z}^{D}$, we can first input minimal state $\mathbf{Z}^{M}_n$ to get the acceleration for integration to get the minimal state at the next time step $\hat{\mathbf{Z}}^{M}_{n+1}$.
 Then, we can solve the dependent state $\hat{\mathbf{Z}}^{D}_{n+1}$ by solving the constraint equation $\hat{\mathbf{Z}}^{D}_{n+1}=\phi^{-1}(\hat{\mathbf{Z}}^{M}_{n+1}, \mathbf{u},\bm{\mu})$. We could use the minimal coordinates $\hat{\mathbf{Z}}^{M}_{n+1}$ and dependent state $\hat{\mathbf{Z}}^{D}_{n+1}$ to get the full combined states $\Tilde{\mathbf{Z}}_{n+1}=(\hat{\mathbf{Z}}^{M}_{n+1},\hat{\mathbf{Z}}^{D}_{n+1})^T\in \mathbb{R}^{2n_z}$. By the difference between the combined predicted state $\Tilde{\mathbf{Z}}_{n+1}$ and the ground truth state $\mathbf{Z}_{n+1}$, we can optimize the \updatedTextRebutalTwo{MBD-NODE}.
A similar concept has been explored in \cite{beucler2021enforcing,daems2022keycld} for enforcing hard constraints within data-driven models. Beucler et al.~\cite{beucler2021enforcing} have approached this by designing a constraint layer, while Daems et al.~\cite{daems2022keycld} encoded holonomic constraints directly into the Euler-Lagrange equations. Our method can address more general non-holonomic constraints.
Given the initial state $\mathbf{Z}_0$ and the ground truth state $\mathbf{Z}_1$, the corresponding loss function for one data pair could be written as:}

\updatedText{
\begin{equation}
    \begin{aligned}
        \text{L}(\mathbf{\Theta}) &= \lVert (\Phi(\mathbf{Z}^{M}_0,f,\Delta t),\phi^{-1}(\Phi(\mathbf{Z}_0^{M},f,\Delta t)))^T - \mathbf{Z}_1\rVert^2_2 \\
        &= \lVert (\hat{\mathbf{Z}}_1^{M},\hat{\mathbf{Z}}_1^{D})^T - {\mathbf{Z}}_1\rVert^2_2 =\lVert \Tilde{\mathbf{Z}}_1- \mathbf{Z}_1 \rVert^2_2,
        \end{aligned}
        \end{equation}
}
\updatedText{(2) If we have access to only the minimal state information$(\mathbf{Z}^{M}_n)$ —for instance, if we prefer not to expend effort in collecting data on dependent coordinates due to potential costs—we can construct and train the \updatedTextRebutalTwo{MBD-NODE} using solely the minimal coordinates $(\mathbf{Z}^{M})$. 
During the inference, we could use \updatedTextRebutalTwo{MBD-NODE} to predict minimal states and then solve all the states. In this case, the loss function could be written as:}
\updatedText{
\begin{equation}
\text{L}(\mathbf{\Theta}) = \lVert \Phi(\mathbf{Z}^{M}_0,f,\Delta t)- \mathbf{Z}^{M}_1\rVert^2_2 =\lVert \hat{\mathbf{Z}}^{M}_1- \mathbf{Z}^{M}_1 \rVert^2_2,
\end{equation}
}

By solving the constraint equation in both the training \updatedText{(with dependent coordinates data)} and inference stage, the hard constraints are satisfied in both phases. \updatedText{The algorithm for constraints equation-based optimization is summarized in the Algorithm \ref{alg:training_hard} (which utilizes dependent coordinates data) and Algorithm \ref{alg:training_hard_minimal} (which uses only minimal coordinates data), both found in Appendix A.}

\subsubsection{Baseline Models}
Table~\textsc{\ref{tab:methods_comparison}} summarizes the models used in the numerical tests discussed in this manuscript, along with some of their salient attributes. Code for all of these methods \updatedText{is} provided with this contribution.

\begin{table}[ht]
\centering
\caption{Summary of the methods comparison. The compared methods are \updatedTextRebutalTwo{MBD-NODE}, HNN, LNN, LSTM, and FCNN.}
\label{tab:methods_comparison}
\begin{tabular}{@{}lccccc@{}} 
\toprule 
& \updatedTextRebutalTwo{MBD-NODE}  & HNN & LNN  & LSTM  & FCNN \\
\midrule 
Works on energy-conserving system & \checkmark & \checkmark & \checkmark &  & \\
Works on general coordinates  & \checkmark&  & \checkmark& \checkmark & \checkmark \\
Works on dissipative systems & \checkmark & &  & \checkmark & \checkmark \\
Works with constraints &\checkmark&&&&\\
No need for second-order derivatives&\checkmark&&\checkmark&& \\
\updatedTextRebutalTwo{Scalability for long time simulation}&\checkmark&\checkmark&\checkmark&&\\
Learn continuous dynamics &\checkmark &\checkmark &\checkmark &  & \checkmark \\

\bottomrule 
\end{tabular}
\end{table}

\section{Numerical Experiments}
\label{sec:numericalexperiments}
We study the performance \updatedText{of} the methods in Table~\ref{tab:methods_comparison} with \updatedText{seven} numerical examples, reflecting on method attributes such as energy conservation, energy dissipation, multi-scale dynamics, \updatedText{generalization to different parameters and external force}, \updatedText{model-based control}, chaotic dynamics, and \updatedText{constraint enforcement.
One or more of these attributes oftentimes comes into play in engineering applications that rely on MBD simulation}. We use these numerical examples to compare the performance of the proposed \updatedTextRebutalTwo{MBD-NODE} methodology with state-of-the-art data-driven modeling methods.
The numerical examples and modeling methods are summarized in Table \ref{tab: tests}.
\begin{table}[htbp]
    \centering
    \caption{Summary of the numerical examples and the modeling methods.}
    \begin{tabular}{@{}lccc@{}} 
    \toprule
        Test Case & \multicolumn{1}{c}{Model A} & \multicolumn{1}{c}{Model B} & \multicolumn{1}{c}{Model C} \\
    \midrule
        Single Mass-Spring & \updatedTextRebutalTwo{MBD-NODE} & HNN & LNN \\
        Single Mass-Spring-Damper & \updatedTextRebutalTwo{MBD-NODE} & LSTM & FCNN \\
        Triple Mass-Spring-Damper & \updatedTextRebutalTwo{MBD-NODE} & LSTM & FCNN \\
        \updatedText{Single Pendulum} & \updatedText{\updatedTextRebutalTwo{MBD-NODE}} & \updatedText{LSTM} & \updatedText{FCNN} \\
        Double Pendulum & \updatedTextRebutalTwo{MBD-NODE} & LSTM & FCNN \\
        \updatedText{Cart-pole} & \updatedText{\updatedTextRebutalTwo{MBD-NODE}} & \updatedText{LSTM} & \updatedText{FCNN} \\
        Slider Crank & \updatedTextRebutalTwo{MBD-NODE} & \updatedText{-} & \updatedText{-} \\
    \bottomrule
    \end{tabular}
    \label{tab: tests}
\end{table}
The model performance is evaluated via the \updatedText{MSE $\epsilon$}, defined by the following equation:

\begin{equation}
    \epsilon = \frac{1}{N} \sum_{i=1}^N \left(\lVert \mathbf{Z}_i - \hat{\mathbf{Z}}_i\rVert^2\right)=\frac{1}{N} \sum_{i=1}^N \left(\lVert \mathbf{z}_i - \hat{\mathbf{z}}_i\rVert^2+\lVert \mathbf{\dot z}_i - \hat{\dot{\mathbf{z}}}_i\rVert^2\right),
\end{equation}
where $i$ indicates the index of a test sample, $\mathbf{z}_i$ and $\mathbf{\dot z}_i$ are the ground truth of the coordinate and its time derivative, while $\hat{\mathbf{z}}_i$ and $\hat{\dot{\mathbf{z}}}_i$ denote the predicted results by a trained model. Here $\|\cdot\|$ corresponds to the standard vector 2-norm.


In Table \ref{tab: model_errors}, we summarize the MSE error made by each method on the test data of all the numerical examples. Sections \ref{sec:single_mass_spring} to \ref{sec:slider_crank} present more details about the setup of each test case and the performance of our method in comparison to the others.
\updatedText{The training cost for the \updatedTextRebutalTwo{MBD-NODE}, HNN, LNN, LSTM, and FCNN models with different integrators used for each test case is recorded in the Appendix \ref{appsec:time_cost}. Python code is provided publicly for all models and all test cases for unfetter used and reproducibility studies \cite{MNODE_supportData2024}.}

\begin{table}[htbp]
    \centering
    \caption{Summary of the numerical error for different models in various test cases. The detailed information about the models is included in Table \ref{tab: tests}.}
    \begin{tabular}{@{}lccc@{}} 
    \toprule
        \multirow{2}{*}{Test Case} & \multicolumn{3}{c}{Error} \\
        \cmidrule(l){2-4} 
        & Model A & Model B & Model C \\
    \midrule
        Single Mass-Spring & 1.3e-6 & 1.9e-2 & 9.1e-6 \\
        Single Mass-Spring-Damper & 8.6e-4 & 1.8e-2  & 9.9e-2  \\
        Triple Mass-Spring-Damper & 8.2e-3 & 1.8e-1  & 4.2e-2 \\
        \updatedText{Single Pendulum} & \updatedText{2.0e-3} & \updatedText{3.4e-3} & \updatedText{8.0e-1}  \\
        Double Pendulum & 2.0e-1 & 6.4e-1 & 2.2e0  \\
        \updatedText{Cart-pole} & \updatedText{6.0e-5} & \updatedText{3.2e-4} & \updatedText{4.7e-2} \\
        Slider Crank & \updatedText{3.2e-2} & \updatedText{-}  & \updatedText{-} \\
    \bottomrule
    \end{tabular}
    \label{tab: model_errors}
\end{table}

\updatedText{\subsection{Single Mass-Spring System}}\label{sec:single_mass_spring}
This system is relevant as it does not model \updatedText{viscous damping} and serves as a numerical example to evaluate the predictive attribute of the trained models on an energy-conserving system \cite{HNN,HNN_IET,NSF_HNN}. 
Figure \ref{fig:single_mass_spring} illustrates the setup of the single mass-spring system. The equation of motion is formulated as
\begin{equation}\label{eqn:eom_sms}
    \frac{d^2 x}{dt^2} = -\frac{k}{m}x ,
\end{equation}
where $x$ represents the displacement of the mass from its equilibrium position; $k$, the spring constant, is set to 50 N/m; and $m$, the mass of the object, is set to 10 kg. The system's Hamiltonian, which describes its total energy, is 
\begin{align}
T(p) &= \frac{p^2}{2m} \\
V(q) &= 1/2kq^2\\
H(p, q) &= T(p)+V(q)\label{eqn:separble}
\end{align}
where: \( q \) is the generalized position; \( p \) is the generalized momentum, which, in this context, is \( m \times \dot{q} \), with \( \dot{q} \) being the generalized velocity; and \(T\) and \(V\) represents the kinetic energy and potential energy.

\begin{figure}[htbp]
    \centering
    \begin{tikzpicture}[x=0.75pt,y=0.75pt,yscale=-1,xscale=1]
        \draw [line width=1.5]    (198,189) -- (436.17,189.5) ;
     
        \draw [line width=1.5]    (198,109.5) -- (198,189) ;
        \draw   (198,164.75) .. controls (203.25,159.13) and (207.25,153.5) .. (215.25,153.5) .. controls (231.25,153.5) and (231.25,176) .. (225.25,176) .. controls (219.25,176) and (219.25,153.5) .. (235.25,153.5) .. controls (251.25,153.5) and (251.25,176) .. (245.25,176) .. controls (239.25,176) and (239.25,153.5) .. (255.25,153.5) .. controls (271.25,153.5) and (271.25,176) .. (265.25,176) .. controls (259.25,176) and (259.25,153.5) .. (275.25,153.5) .. controls (291.25,153.5) and (291.25,176) .. (285.25,176) .. controls (279.25,176) and (279.25,153.5) .. (295.25,153.5) .. controls (297.54,153.5) and (299.5,153.96) .. (301.17,154.75) ;
        \draw   (301.17,126.5) -- (369,126.5) -- (369,189) -- (301.17,189) -- cycle ;

        \draw [dashed][->,line width=1.5] (198,189) -- (470,189) node[anchor=north] {$x$};

        \draw (326,148) node [anchor=north west][inner sep=0.75pt]   [align=left] {$\displaystyle \mathbf{m}$};
        \draw (247,130) node [anchor=north west][inner sep=0.75pt]   [align=left] {$\displaystyle \mathbf{k}$};
    \end{tikzpicture}
    \caption{Single mass-spring system; $k$ and $m$ denote the spring constant and the mass of the object, respectively. Only the motion along $x$-direction is considered.}
    \label{fig:single_mass_spring}
\end{figure}
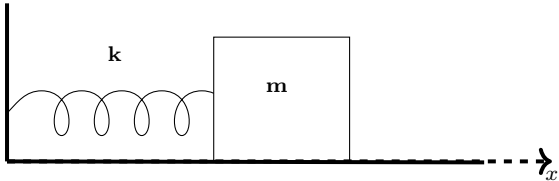
We choose a time step of 0.01s in both training and testing for the single mass-spring system. The training data consists of a trajectory analytically solved over 300 \updatedText{time steps} with initial conditions $x_0=1\ \unit{m}, v_0=0\ \unit{m/s}$.

For this system, which possesses a separable Hamiltonian as shown in Eq.~\eqref{eqn:separble}, the \updatedTextRebutalTwo{MBD-NODE} model employed the leapfrog method as the symplectic integrator of choice. We also show the performance of \updatedTextRebutalTwo{MBD-NODE} when used with the more common RK4 \updatedText{integrator}. We also benchmark against the HNN and LNN methods (see Table~\ref{tab: tests} for a summary). The Hamiltonian-based methods used data in generalized coordinates, while the others (including a numerical method) were tested using Cartesian coordinates. We also provide a baseline test by numerically solving the system of ODEs in Eq.~\eqref{eqn:eom_sms} with the RK4 integrator. The specific configurations of each model, including the choice of coordinate systems and integrators, are detailed in Table \ref{tab:sms_test}. Additionally, the hyperparameters used for the neural network-based tests are summarized in Table \ref{tab:hyper_sms}. These settings and tests were designed to evaluate the efficiency and accuracy of different modeling approaches and integrators in predicting and understanding the dynamics of the single mass-spring system.

\begin{table}[htbp]
\centering
\caption{Numerical tests with corresponding MSE for the single-mass spring system}
\label{tab:sms_test}
\begin{tabular}{@{}lcccc@{}} 
\toprule
Model & {Coordinate System} & {Integrator} & MSE \\
\midrule
\updatedTextRebutalTwo{MBD-NODE} & Generalized & {\updatedText{Leapfrog}} & 1.3e-6 \\
HNN & Generalized & {RK4} &2.0e-3\\
LNN & Cartesian & {RK4} &9.1e-6\\
Numerical & Cartesian & {RK4} &2.0e-3\\
\updatedTextRebutalTwo{MBD-NODE} & Cartesian & {RK4} &9.2e-1\\
\bottomrule
\end{tabular}
\end{table}


\begin{table}[htbp]
\centering
\caption{Hyper-parameters for the single mass-spring system}
\label{tab:hyper_sms}
\begin{tabular}{@{}lcccc@{}}
\toprule
Hyper-parameters & \multicolumn{4}{c}{Model} \\
\cmidrule(r){2-5}
                  & \updatedTextRebutalTwo{MBD-NODE}\textsubscript{LF}          &\updatedTextRebutalTwo{MBD-NODE}\textsubscript{RK4}& HNN & LNN \\ \midrule
No. of hidden layers     & \updatedText{2}           & \updatedText{2}                           & \updatedText{2}                        &\updatedText{2}\\
\updatedText{No. of nodes per hidden layer} & \updatedText{256} & \updatedText{256} & \updatedText{256}&\updatedText{256}  \\
Max. epochs                  & 450       & 300                     & 30000&400                  \\
Initial learning rate & 1e-3 & 1e-3 & 1e-3 & 1e-4 \\
Learning rate decay & 0.99  &0.98    &  0.98&0.98       \\ 
Activation function & Tanh & Tanh & Sigmoid,Tanh & Softmax\\
Loss function                  & MSE     & MSE& MSE& MSE                  \\
Optimizer                       & Adam    & Adam & Adam & Adam                                \\  \bottomrule
\end{tabular}
\end{table}

Figure \ref{fig:sms_x_V_time} shows the dynamic response in terms of position $x$ and velocity $v$ for the test data, and the MSE of each method \updatedText{is} shown in Table \ref{tab:sms_test}. 
More specifically, Figs. \ref{fig:sms_x_V_time} (a) to (d) demonstrate the performance of the \updatedTextRebutalTwo{MBD-NODE} model with the RK4 integrator, as well as the results obtained from a purely numerical solution using the RK4 method. 
The ground truth for comparison is obtained by analytically solving Eq.~\eqref{eqn:eom_sms}. It can be seen in Figs. \ref{fig:sms_x_V_time} (c) and (d) that the direct usage of the RK4 integrator provides results that gradually deviate from the true system. 
This issue of gradually increased errors becomes more severe in the \updatedTextRebutalTwo{MBD-NODE} results with the RK4 integrator in Figs. \ref{fig:sms_x_V_time} (a) and (b), highlighting their limitations in accurately modeling Hamiltonian systems.
In contrast, both the LNN and HNN models, despite utilizing the RK4 integrator, demonstrate stable behavior in solving the mass-spring system, as shown in Figs. \ref{fig:sms_x_V_time} (g) to (j). The more stable simulations of these two models can be attributed to the underlying equations of these models, which ensure energy conservation in the system. Notably, the HNN's performance, as shown in Figs. \ref{fig:sms_x_V_time}(i) and (j), show a deviation from the expected trajectory around the 30-second mark, leading to a higher MSE when compared to the \updatedTextRebutalTwo{MBD-NODE} with the leapfrog integrator and the \updatedText{LNN} model. Among all the methods that we studied, the \updatedTextRebutalTwo{MBD-NODE} with the leapfrog integrator outperforms other models, achieving the lowest MSE of $\epsilon$=1.3e-6, with detailed trajectories of $x$ and $v$ presented in Figs. \ref{fig:sms_x_V_time}(e) and (f).
\begin{figure}[htbp]
    \centering
    \includegraphics[width=12cm]{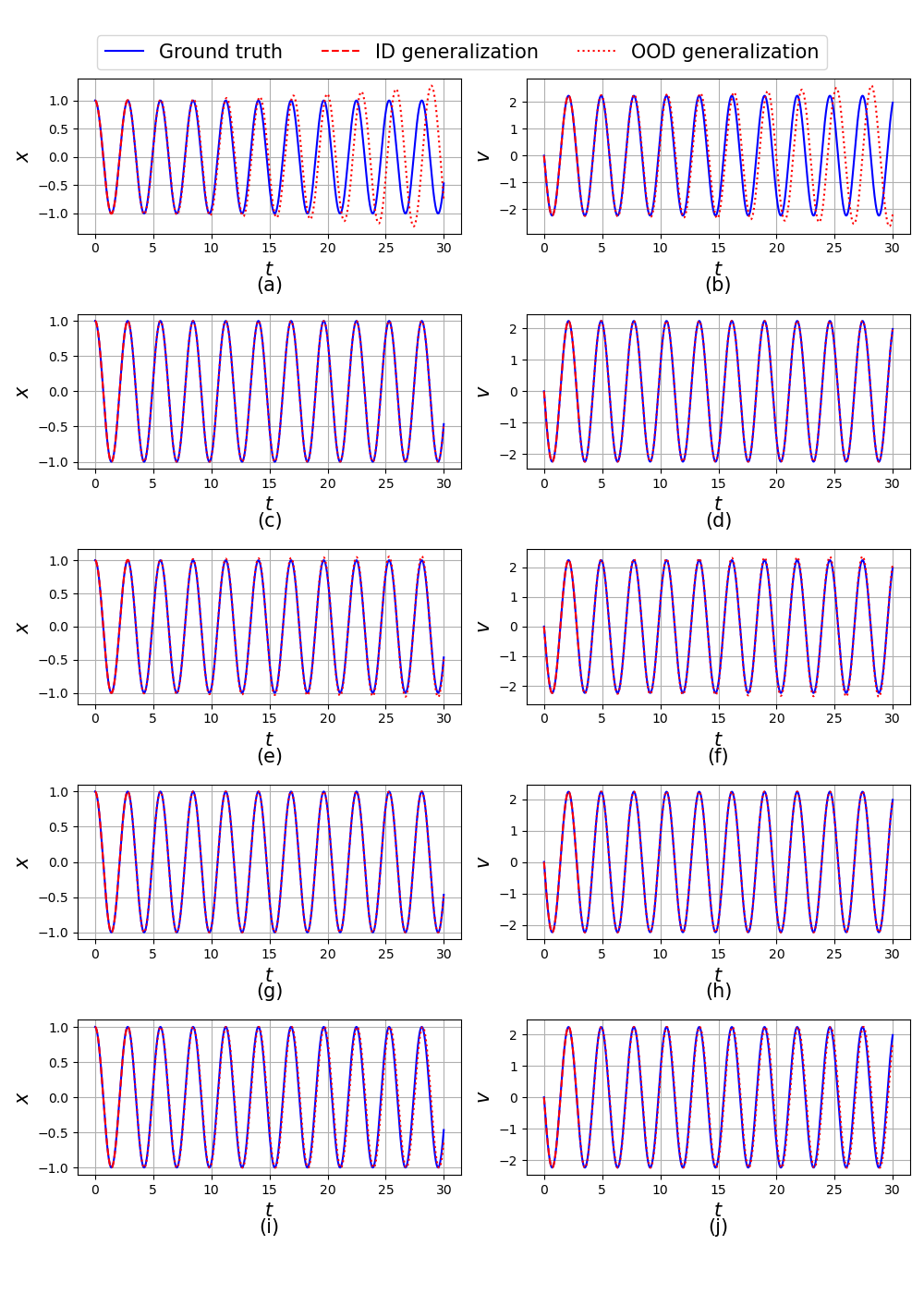}
    \caption{Comparison of $t$ vs $x$ (Left Column) and $t$ vs $v$ (Right Column) for the different model and integrator combinations (rows) for the single-mass-spring system.  Notice the dashed lines represent performance on the training data set $t \in [0,3]$, after which the dotted lines represent performance on the testing data set $t \in [0,30]$. (a), (b) are for the \updatedTextRebutalTwo{MBD-NODE} with RK4; (c), (d) \updatedText{are for the \updatedTextRebutalTwo{MBD-NODE} with leapfrog integrator;} (e), (f) \updatedText{are for the RK4 integrator;} (g), (h) are for the LNN; (i), (j) are for the HNN.}
    \label{fig:sms_x_V_time}
\end{figure}
\begin{figure}[htbp]
    \centering
    \includegraphics[width=12cm]{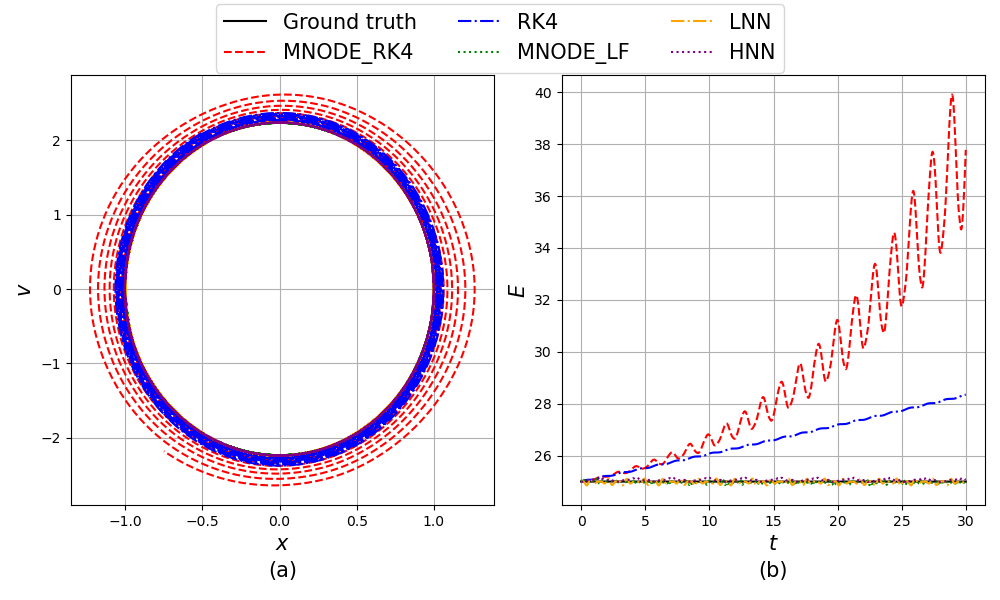}
    \caption{(a) The phase space $x$ vs $v$ and (b) the system energy for the test data for the single mass-spring system}
    \label{fig:sms_phase_energy}
\end{figure}

Figure \ref{fig:sms_phase_energy} presents the phase space trajectory and energy profile for the test set. It confirms the instability issues with the RK4 solver and the \updatedTextRebutalTwo{MBD-NODE} model with the RK4 integrator, particularly in terms of energy drift accumulating over time. In comparison, both the LNN and HNN models, as well as the \updatedTextRebutalTwo{MBD-NODE} model with the leapfrog integrator, demonstrate stable solutions without any noticeable energy drift. The results in Figs. \ref{fig:sms_x_V_time} and \ref{fig:sms_phase_energy} confirm the effectiveness of the \updatedTextRebutalTwo{MBD-NODE} model with a symplectic integrator in accurately learning the Hamiltonian structure of the system.

\updatedText{\subsection{Single Mass-Spring-Damper System}}\label{sec:smsd}
The second numerical test involves a single-mass-spring-damper system, as shown in Fig. \ref{fig:smsd_figure}. Compared with the first numerical test, there is a damper between the mass and the wall that causes the mass to slow down over time. It is important to note that the models designed for energy-conserving systems, like the LNN and HNN, are generally not applicable for dissipative systems without further modification. Therefore, we compare the performance of our method to LSTM and FCNN models, which are commonly employed in multibody dynamics problems.

The equation of motion for the single-mass-spring-damper system is given by:
\begin{equation}
\frac{d^2 x}{dt^2} = \frac{-k}{m}x - \frac{d}{m}\frac{dx}{dt},
\end{equation}
where $x$ represents the displacement of the mass from its equilibrium position; $m$ is the mass of the object, set to 10 $\unit{kg}$ in this test; $d$ is the damping coefficient, set to $2\ \unit{Ns/m}$ in this test; and $k$ is the coefficient of stiffness of the spring, set to $50\ \unit{N/m}$.
\begin{figure}[htbp]
    \centering
    \begin{tikzpicture}[x=0.75pt,y=0.75pt,yscale=-1,xscale=1]
        \draw [line width=1.5]    (198,189) -- (436.17,189.5) ;
        \draw [line width=1.5]    (201.17,98.5) -- (202,189) ;
        \draw   (200.42,120) .. controls (214.42,120) and (214.42,141) .. (210.42,141) .. controls (206.42,141) and (206.42,120) .. (220.42,120) .. controls (234.42,120) and (234.42,141) .. (230.42,141) .. controls (226.42,141) and (226.42,120) .. (240.42,120) .. controls (254.42,120) and (254.42,141) .. (250.42,141) .. controls (246.42,141) and (246.42,120) .. (260.42,120) .. controls (274.42,120) and (274.42,141) .. (270.42,141) .. controls (266.42,141) and (266.42,120) .. (280.42,120) .. controls (294.42,120) and (294.42,141) .. (290.42,141) .. controls (286.46,141) and (286.42,120.42) .. (300,120.01) ;
        \draw   (301.17,111.5) -- (369,111.5) -- (369,189) -- (301.17,189) -- cycle ;
        \draw    (240.23,164) -- (240.23,176) ;
        \draw    (201,170) -- (239.61,169.51) ;
        \draw    (259.41,170) -- (300,170) ;
        \draw   (239.61,160) -- (259.41,160) -- (259.41,180) ;
        \draw   (239.61,180) -- (259.41,180) -- (259.41,168.51) ;

        \draw [dashed][->,line width=1.5] (198,189) -- (470,189) node[anchor=north] {$x$};

        \draw (327,142) node [anchor=north west][inner sep=0.75pt]   [align=left] {$\displaystyle \mathbf{m}$};
        \draw (261,102) node [anchor=north west][inner sep=0.75pt]  [font=\small] [align=left] {$\displaystyle \mathbf{m}$};
        \draw (261,152) node [anchor=north west][inner sep=0.75pt]  [font=\small] [align=left] {$\displaystyle \mathbf{d}$};

    \end{tikzpicture}
    \caption{The single mass-spring-damper system. The setup is similar to the example in Fig. \ref{fig:single_mass_spring}, except for the addition of a damper with coefficient $d$ here.}
    \label{fig:smsd_figure}
\end{figure}
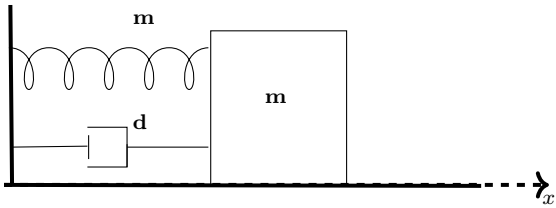

We choose the time step as 0.01s for both the training and testing. The training dataset consists of a trajectory numerically solved by the RK4 solver for 300 time steps. In the testing phase, the models are tested by predicting the system state within the training range and extrapolation to predict system behavior for an additional \updatedText{100 time steps}. The initial condition for this problem is \updatedText{$x=1\ \unit{m}, v=0\ \unit{m/s}$}. It should be noted that the system is no longer a Hamiltonian system, so we use Cartesian coordinates for all the methods. The hyperparameters used for each model are summarized in Table \ref{tab:hyper_smsd}.

\begin{table}[htbp]
\centering
\caption{Hyper-parameters for the single mass-spring-damper system.}
\label{tab:hyper_smsd}
\begin{tabular}{@{}lcccc@{}}
\toprule
Hyper-parameters & \multicolumn{3}{c}{Model} \\
\cmidrule(r){2-4}
                  & \updatedTextRebutalTwo{MBD-NODE}          &LSTM & FCNN \\ \midrule
No. of hidden layers     & \updatedText{2}            & \updatedText{2}                            & \updatedText{2}                         \\
\updatedText{No. of nodes per hidden layer}  & \updatedText{256}  & \updatedText{256} & \updatedText{256}  \\

Max. epochs                  & 350       & 400                     & 600                  \\
Initial learning rate &  1e-3 &5e-4 & 5e-4 \\
Learning rate decay & 0.98  &0.98    &  0.98       \\ 
Activation function & Tanh &  Sigmoid,Tanh &Tanh\\
Loss function                  & MSE   & MSE & MSE                    \\
Optimizer                       & Adam & Adam & Adam                                       \\  \bottomrule
\end{tabular}
\end{table}

Figure \ref{fig:smsd_xvt} presents the position $x$ and velocity $v$ for all the trained models. In the first 300 time steps, which correspond to the training range, all three models exhibit accurate predictions, indicating an effective training process. However, differences in model performance start to show up in the testing regime (i.e., $t>3$). More specifically, Fig. \ref{fig:smsd_xvt}(a) and (b) show that the \updatedTextRebutalTwo{MBD-NODE} gives a reasonable prediction that closely matches the ground truth with the lowest MSE of $\epsilon=$8.6e-4, which demonstrates its predictive capability.

On the other hand, the LSTM predictions tend to just replicate historical data patterns (see Fig. \ref{fig:smsd_xvt}(c) and (d)), rather than learning and adapting to the underlying dynamics of the system. This limitation makes LSTM fail to correctly capture the decay of energy for an energy-dissipative system. The FCNN model struggles with extrapolation as well, mainly because the good extrapolation performance of FCNN heavily relies on the closeness of training and testing data in their distributions. This limitation of FCNN leads to errors in the extrapolation task of this example, resulting in the largest MSE of $\epsilon= 9.9e-2$ as shown in Fig. \ref{fig:smsd_xvt}(e) and (f).

\begin{figure}[htbp]
    \centering
    \includegraphics[width=12cm]{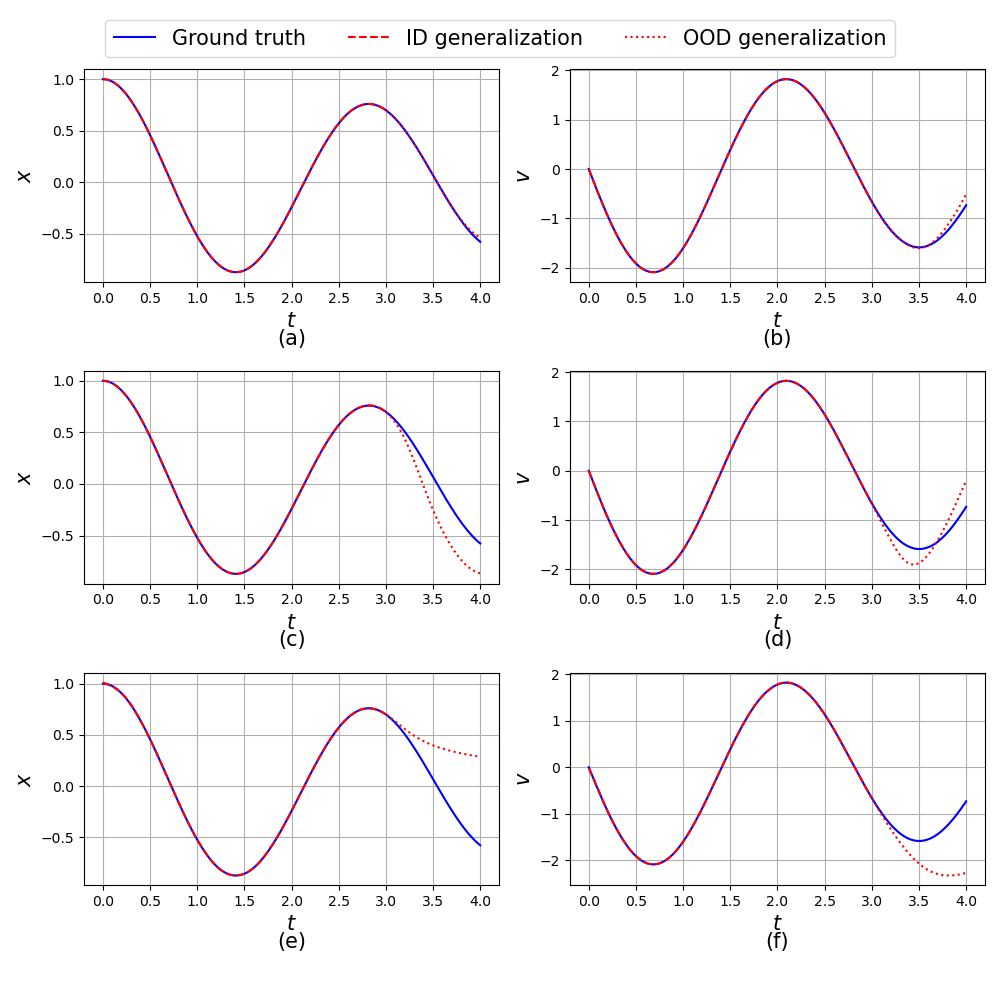}
    \caption{Comparison of $t$ vs $x$ (Left Column) and $t$ vs $v$ (Right Column) for the different models (rows) for the single-mass-spring-damper system. Notice the dashed lines represent performance on the training data set $t \in [0,3]$ after which the dotted lines represent performance on the testing data set. (a), (b) are for the \updatedTextRebutalTwo{MBD-NODE} with MSE $\epsilon=$8.6e-4; (c), (d) are for the LSTM with MSE $\epsilon=$1.8e-2; (e), (f) are for the FCNN with MSE $\epsilon=$ 9.9e-2.}
    \label{fig:smsd_xvt}
\end{figure}

More insights into the system's dynamics \updatedText{are} provided by the phase space trajectories illustrated in Fig. \ref{fig:smsd_phase}. The predictions of the \updatedTextRebutalTwo{MBD-NODE}, as depicted in Fig. \ref{fig:smsd_phase}(a), are closely aligned with the observed behavior of the system. In contrast, the LSTM's trajectory, shown in Fig. \ref{fig:smsd_phase}(b), exhibits stagnation and fails to reflect the system's eventual halt. The FCNN's performance, presented in Fig. \ref{fig:smsd_phase}(c), is lacking during the extrapolation test -- it merely yields predictions in the tangent direction, resulting in a significant divergence from the anticipated trajectory.


\begin{figure}[htbp]
    \centering
    \includegraphics[width=12cm]{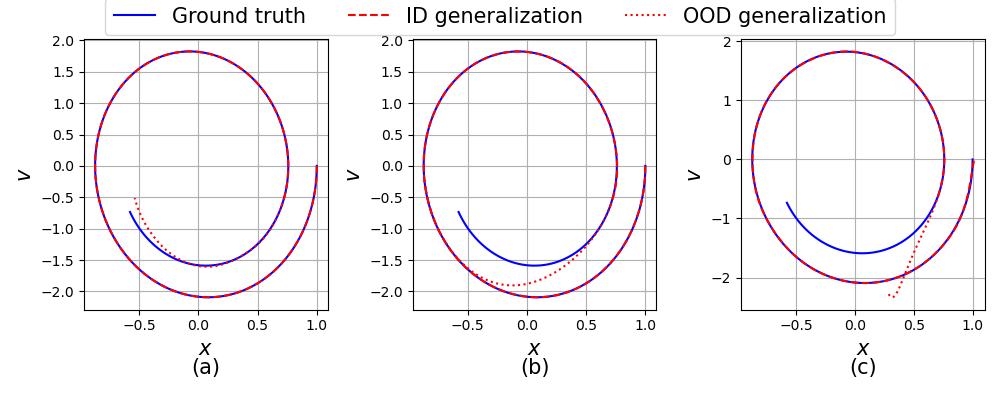}
    \caption{The phase space $x$ vs $v$ for the single mass spring damper system. Dashed lines represent performance on the training data, and the dotted lines on the test data. (a) is for the \updatedTextRebutalTwo{MBD-NODE}; (b) is for the LSTM; (c) is for the FCNN.}
    \label{fig:smsd_phase}
\end{figure}

\subsection{Multiscale Triple Mass-Spring-Damper System}

This system, shown in Fig. \ref{fig:tmsd_figure}, has three masses. The largest mass is 100 times larger than the smallest one. The main purpose of this example is to gauge method performance on multiscale systems. The equations of motion for the triple mass-spring-damper system are as follows:
\begin{equation}
\begin{aligned}
\frac{d^2x_1}{dt^2} &= -\frac{k_1}{m_1} x_1 - \frac{d_1}{m_1} (v_1 - v_2) + \frac{k_2}{m_1} (x_2 - x_1) + \frac{d_2}{m_1} (v_2 - v_1),\\
\frac{d^2x_2}{dt^2} &= -\frac{k_2}{m_2} (x_2 - x_1) - \frac{d_2}{m_2} (v_2 - v_1) + \frac{k_3}{m_2} (x_3 - x_2) + \frac{d_3}{m_2} (v_3 - v_2), \\
\frac{d^2x_3}{dt^2} &= -\frac{k_3}{m_3} (x_3 - x_2) - \frac{d_3}{m_3} (v_3 - v_2) ,\\
\end{aligned}
\end{equation}
where $x_1,x_2,x_3$ are the positions of the masses, respectively; $m_1,m_2,m_3$ are the masses of the object with values of 100 kg, 10 kg, and 1 kg, respectively; $d_1,d_2,d_3$ are the damping coefficients, each set to 2 Ns/m; and $k_1,k_2,k_3$ are the spring stiffness values, all set to 50 N/m.
\begin{figure}[htbp]
    \centering
    \begin{tikzpicture}[x=0.75pt,y=0.75pt,yscale=-0.75,xscale=0.75]
    
    \draw [line width=1.5]    (49.67,209) -- (564.17,209.5) ;
   
    \draw [line width=1.5]    (55.08,118.5) -- (56.5,209) ;
    \draw   (56.99,141.5) .. controls (68.64,141.5) and (68.64,162.5) .. (65.31,162.5) .. controls (61.98,162.5) and (61.98,141.5) .. (73.64,141.5) .. controls (85.29,141.5) and (85.29,162.5) .. (81.96,162.5) .. controls (78.63,162.5) and (78.63,141.5) .. (90.29,141.5) .. controls (101.94,141.5) and (101.94,162.5) .. (98.61,162.5) .. controls (95.28,162.5) and (95.28,141.5) .. (106.94,141.5) .. controls (118.59,141.5) and (118.59,162.5) .. (115.26,162.5) .. controls (111.93,162.5) and (111.93,141.5) .. (123.59,141.5) .. controls (135.24,141.5) and (135.24,162.5) .. (131.91,162.5) .. controls (128.62,162.5) and (128.58,141.92) .. (139.89,141.51) ;
    \draw   (140.72,131.5) -- (203.77,131.5) -- (203.77,209) -- (140.72,209) -- cycle ;
    \draw   (89.04,180) -- (105.98,180) -- (105.98,200) ;
    \draw   (89.04,200) -- (105.98,200) -- (105.98,188.51) ;
    \draw    (89.77,184) -- (89.77,196) ;
    \draw    (56,190) -- (89.04,189.51) ;
    \draw    (105.98,190) -- (140.72,190) ;
    \draw   (203.22,142.5) .. controls (214.87,142.5) and (214.87,163.5) .. (211.54,163.5) .. controls (208.21,163.5) and (208.21,142.5) .. (219.87,142.5) .. controls (231.52,142.5) and (231.52,163.5) .. (228.19,163.5) .. controls (224.86,163.5) and (224.86,142.5) .. (236.52,142.5) .. controls (248.17,142.5) and (248.17,163.5) .. (244.84,163.5) .. controls (241.51,163.5) and (241.51,142.5) .. (253.17,142.5) .. controls (264.82,142.5) and (264.82,163.5) .. (261.49,163.5) .. controls (258.16,163.5) and (258.16,142.5) .. (269.82,142.5) .. controls (281.47,142.5) and (281.47,163.5) .. (278.14,163.5) .. controls (274.85,163.5) and (274.81,142.92) .. (286.12,142.51) ;
    \draw   (286.95,132.5) -- (350,132.5) -- (350,210) -- (286.95,210) -- cycle ;
    \draw   (235.27,181) -- (252.21,181) -- (252.21,201) ;
    \draw   (235.27,201) -- (252.21,201) -- (252.21,189.51) ;
    \draw    (236,185) -- (236,197) ;
    \draw    (202.23,191) -- (235.27,190.51) ;
    \draw    (252.21,191) -- (286.95,191) ;
    \draw   (350.99,142.5) .. controls (362.64,142.5) and (362.64,163.5) .. (359.31,163.5) .. controls (355.98,163.5) and (355.98,142.5) .. (367.64,142.5) .. controls (379.29,142.5) and (379.29,163.5) .. (375.96,163.5) .. controls (372.63,163.5) and (372.63,142.5) .. (384.29,142.5) .. controls (395.94,142.5) and (395.94,163.5) .. (392.61,163.5) .. controls (389.28,163.5) and (389.28,142.5) .. (400.94,142.5) .. controls (412.59,142.5) and (412.59,163.5) .. (409.26,163.5) .. controls (405.93,163.5) and (405.93,142.5) .. (417.59,142.5) .. controls (429.24,142.5) and (429.24,163.5) .. (425.91,163.5) .. controls (422.62,163.5) and (422.58,142.92) .. (433.89,142.51) ;
    \draw   (434.72,132.5) -- (497.77,132.5) -- (497.77,210) -- (434.72,210) -- cycle ;
    \draw   (383.04,181) -- (399.98,181) -- (399.98,201) ;
    \draw   (383.04,201) -- (399.98,201) -- (399.98,189.51) ;
    \draw    (383.77,185) -- (383.77,197) ;
    \draw    (350,191) -- (383.04,190.51) ;
    \draw    (399.98,191) -- (434.72,191) ;

    \draw (162.1,163) node [anchor=north west][inner sep=0.75pt]   [align=left] {$\displaystyle \mathbf{m_{1}}$};
    \draw (105.45,122) node [anchor=north west][inner sep=0.75pt]  [font=\small] [align=left] {$\displaystyle \mathbf{k_{1}}$};
    \draw [dashed][->,line width=1.5] (49.67,209) -- (600,209.5) node[anchor=north] {$\displaystyle \mathbf{x}$};
  
    \draw (107,172) node [anchor=north west][inner sep=0.75pt]  [font=\small] [align=left] {$\displaystyle \mathbf{d_{1}}$};
    \draw (308.33,164) node [anchor=north west][inner sep=0.75pt]   [align=left] {$\displaystyle \mathbf{m_{2}}$};
    \draw (251.68,123) node [anchor=north west][inner sep=0.75pt]  [font=\small] [align=left] {$\displaystyle \mathbf{k_{2}}$};
    \draw (253.71,173) node [anchor=north west][inner sep=0.75pt]  [font=\small] [align=left] {$\displaystyle \mathbf{d_{2}}$};
    \draw (456.1,164) node [anchor=north west][inner sep=0.75pt]   [align=left] {$\displaystyle \mathbf{m_{3}}$};
    \draw (399.45,123) node [anchor=north west][inner sep=0.75pt]  [font=\small] [align=left] {$\displaystyle \mathbf{k_{3}}$};
    \draw (401.48,173) node [anchor=north west][inner sep=0.75pt]  [font=\small] [align=left] {$\displaystyle \mathbf{d_{3}}$};
    \end{tikzpicture}
    \caption{\updatedText{Triple mass-spring-damper system. The setup is similar to the example in Fig. \ref{fig:single_mass_spring}, except for the addition of two more masses, springs, and dampers.}}
    \label{fig:tmsd_figure}
\end{figure}
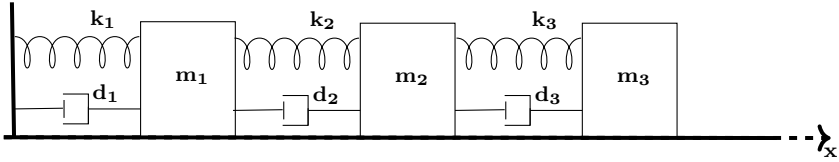

For the numerical settings of the triple mass-spring-damper system, we choose the time step as 0.01s for both training and testing. The training dataset has a trajectory numerically computed by the RK4 solver for \updatedText{300 time steps}. The initial conditions are set as $x_1=1,x_2=2,x_3=3,v_1=v_2=v_3=0$ (all units are SI). The models are tested by extrapolating for 100 more time steps. The hyperparameters used for the models are summarized in Table \ref{tab:hyper_tmsd}.

\begin{table}[h]
\centering
\caption{Hyper-parameters for the triple mass-spring-damper system. }
\label{tab:hyper_tmsd}
\begin{tabular}{@{}lcccc@{}}
\toprule
Hyper-parameters & \multicolumn{3}{c}{Model} \\
\cmidrule(r){2-4}
                  & \updatedTextRebutalTwo{MBD-NODE}          &LSTM & FCNN \\ \midrule
No. of hidden layers    & \updatedText{2}          & \updatedText{2}                           & \updatedText{2}                        \\
\updatedText{No. of nodes per hidden layer}  & \updatedText{256}& \updatedText{256} & \updatedText{256}  \\
Max. epochs                  & 350       & 400                     & 600                  \\
Initial learning rate &  6e-4 &5e-4 & 5e-4 \\
Learning rate decay & 0.98  &0.98    &  0.98       \\ 
Activation function & Tanh &  Sigmoid,Tanh &Tanh\\
Loss function                  & MSE   & MSE& MSE                    \\
Optimizer                       & Adam      & Adam  & Adam                                \\  \bottomrule
\end{tabular}
\end{table}

Figure \ref{fig:tmsd_xvt} presents the position $x$ and velocity $v$ of the triple mass-spring-damper system during training and testing. In terms of accuracy, the \updatedTextRebutalTwo{MBD-NODE} outperforms other models with an MSE $\epsilon$= 8.2e-3. More specifically, the \updatedTextRebutalTwo{MBD-NODE} and LSTM models provide accurate results in the range of training data (i.e., $t<3$) while the results of FCNN model show small oscillation mainly due to the \updatedText{multiscale setting} of the dynamics shown in Fig. \ref{fig:tmsd_xvt}(e). In the testing data (i.e., $t>3$), the performance of trained models starts to differ more. The \updatedTextRebutalTwo{MBD-NODE} can still give a reasonable prediction for the triple mass-spring-damper system shown in Fig. \ref{fig:tmsd_xvt}(a) and (b), although the predicted trajectories slowly deviate from the true ones, mainly because of the accumulation of numerical errors. On the other hand, the LSTM tends to replicate some of the historical patterns. The testing performance of the FCNN model is more reasonable than the LSTM model in this example, while still less satisfactory compared with the \updatedTextRebutalTwo{MBD-NODE} model.

\begin{figure}[htbp]
    \centering
    \includegraphics[width=12cm]{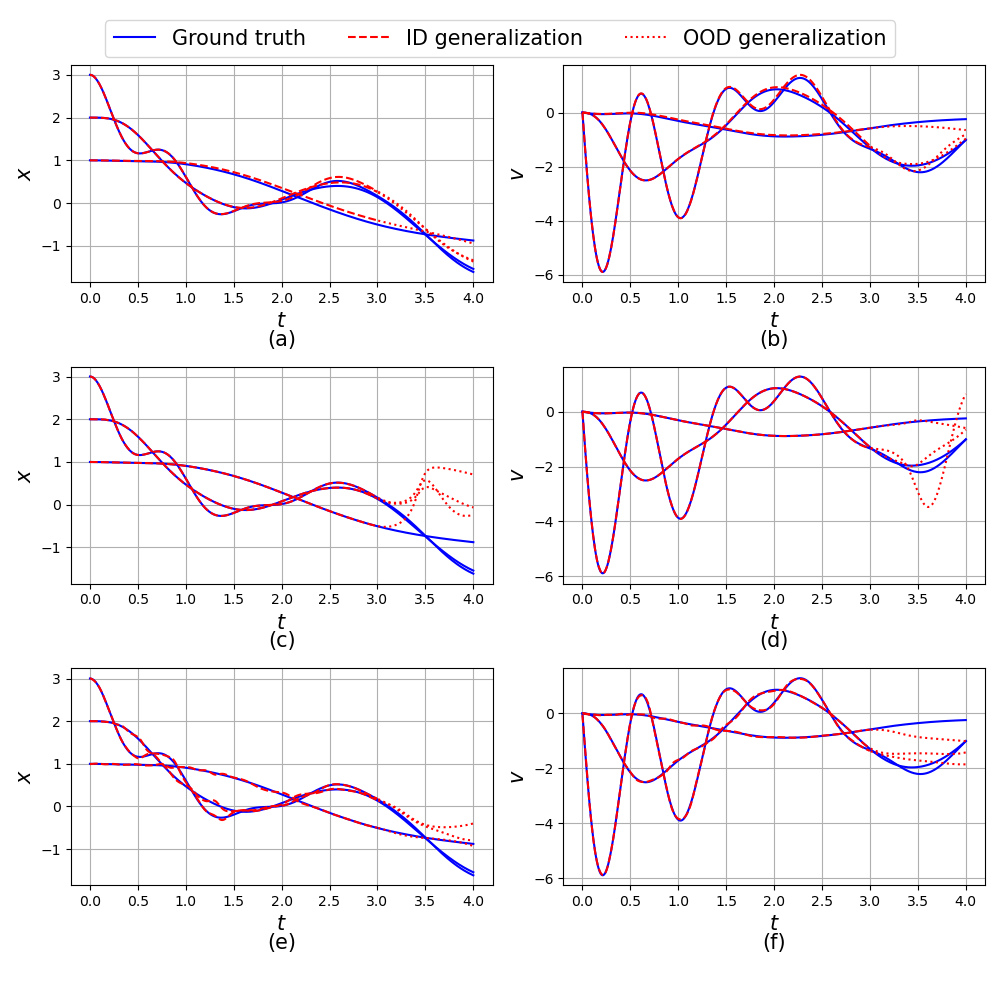}
    \caption{Comparison of $t$ vs $x$ (Left Column) and $t$ vs $v$ (Right Column) for the different models (rows) for the triple-mass-spring-damper system. Notice the dashed lines represent performance on the training data set $t \in [0,3]$ after which the dotted lines represent performance on the testing data set. (a), (b) are for the \updatedTextRebutalTwo{MBD-NODE} with MSE $\epsilon=$8.2e-3; (c), (d) are for the LSTM with MSE $\epsilon=$1.8e-2; (e), (f) are for the FCNN with MSE $\epsilon=$ 4.2e-2.}
    \label{fig:tmsd_xvt}
\end{figure}

The trajectory for the triple mass spring damper system for the test set is shown in Fig. \ref{fig:tmsd_phase}. We can see that for the \updatedTextRebutalTwo{MBD-NODE}, the trajectory of the first body shown in Fig. \ref{fig:tmsd_phase}(a) has some mismatch with the ground truth. This is caused by the multiscale property that the first mass has the largest mass which leads to the \updatedText{slightest} change in the position $x$ and velocity $v$, while the \updatedTextRebutalTwo{MBD-NODE} learns the dynamics from the difference between the state at two nearby times. So, the largest mass will contribute the least to the loss, which causes the \updatedTextRebutalTwo{MBD-NODE} to learn the dynamics of the first body \updatedText{inadequately}. The numerical integration error also \updatedText{accumulates} during the inference, which makes the error larger. For LSTM, we can more clearly see its prediction trends converge to the historical data, which does not work well during \updatedText{OOD generalization}. For FCNN, we note the oscillation for the first body in \updatedText{ID generalization}, and for the \updatedText{OOD generalization}, which leads to a lackluster predictive performance.
\begin{figure}[htbp]
    \centering
    \includegraphics[width=12cm]{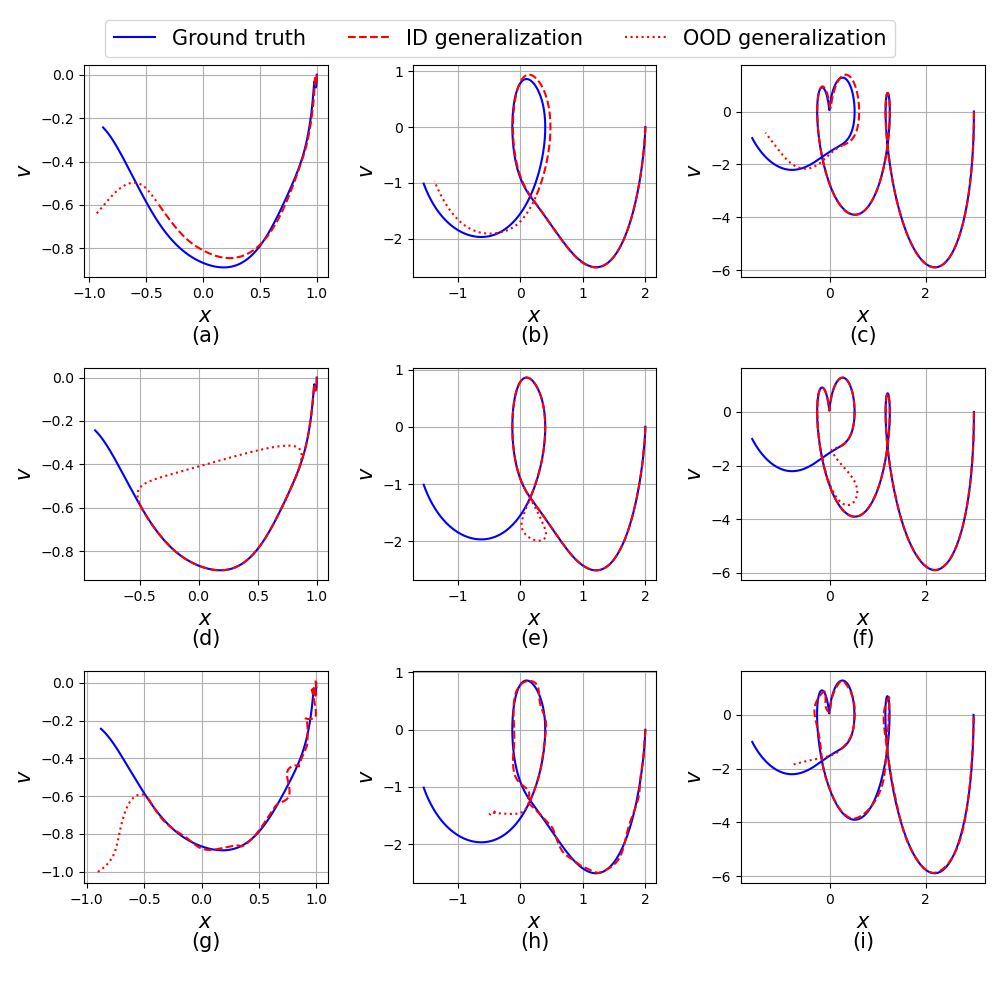}
    \caption{The phase space trajectories for the triple mass spring damper system. The left, middle, and right columns correspond to the first, second, and third mass. Dashed lines represent performance on the training data, and the dotted lines on the test data.\updatedText{(a), (b), (c) are for the \updatedTextRebutalTwo{MBD-NODE}; (d), (e), (f) are for the LSTM; (g), (h), (i) are for the FCNN.}}
    \label{fig:tmsd_phase}
\end{figure}

\subsection{Damped Single Pendulum}
\updatedText{
In this section, we test the \updatedTextRebutalTwo{MBD-NODE}'s ability to generalization on different initial conditions and external forces using the damped single pendulum as shown in Fig. \ref{fig:single_pendulum}.
The equation of motion Eq.~\eqref{eqn:sp} for a damped single pendulum, including the gravitational and damping forces, can be represented as a second-order ODE as follows:}

\begin{equation}
\ddot{\theta}(t) + \frac{g}{L} \sin(\theta(t)) + \frac{c}{mL} \dot{\theta}(t) = 0,\\
\label{eqn:sp}
\end{equation}

\updatedText{
where $\theta(t)$ is the angular displacement as a function of time, $g$ is the acceleration due to gravity and external force, $L=1$ m is the length of the pendulum, $c=0.1$ Ns/m is the damping coefficient, and $m=1$ kg is the mass of the pendulum bob.}
    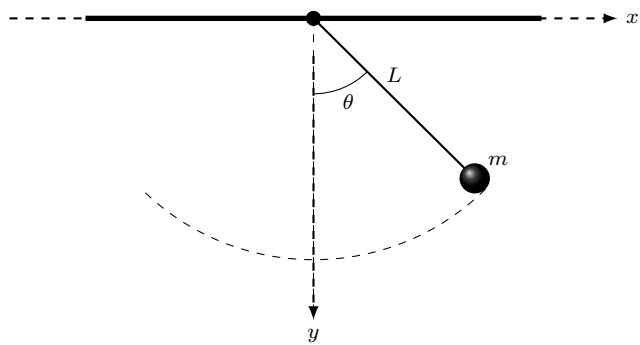
\begin{figure}
    \centering
        \begin{tikzpicture}[>=latex]
            \def\rodlength{3.0}
            \def\massradius{0.2}
            \def\pivotradius{0.1}
            \def\gravitylength{1.0}
            \def\tensionlength{0.5}
            \def\angle{45} 
            \def\axissize{4.0} 
    
            \draw[line width=2] (-\axissize+1,0) -- (\axissize-1,0) ;
            \draw[->,thick,dashed] (-\axissize,0) -- (\axissize,0) node[right] {$x$};
            \draw[->,thick,dashed] (0,-0.5) -- (0,-\rodlength-\gravitylength) node[below] {$y$};
    
            \fill (0,0) circle (\pivotradius) node[above] {};
    
            \draw[thick] (0,0) -- ++(-\angle:\rodlength) node[midway,above=0.1] {$L$};
    
            \shade[ball color=black] (-\angle:\rodlength) circle (\massradius) node[above right=0.1] {$m$};
    
        \draw (0,-1) arc (270:270+\angle:1);
        \node at (270+\angle*0.5:1.2) {$\theta$};
            \draw[dashed] (0,0) -- (0,-\rodlength-\gravitylength);
            \draw[dashed] (0,0) ++(-\angle:\rodlength+\massradius) arc (-\angle:-180+\angle:\rodlength+\massradius);
            
        \end{tikzpicture}
        \caption{\updatedText{Single pendulum system.}}
        \label{fig:single_pendulum}
    \end{figure}

    \updatedText{
Initially, we examine a scenario where the pendulum is released from its lowest point with an initial angular velocity of $\omega = \pi$. We employ various models to predict the trajectory of the pendulum using identical training and testing datasets. In practice, the midpoint method is utilized to solve the ODE Eq.~\eqref{eqn:sp}, adopting a time step of 0.01 seconds. The dataset for training spans the initial 3 seconds, whereas the testing dataset covers the subsequent 1 second. The hyperparameters applied across the models are detailed in Table \ref{tab:hyper_sp}.
}

\begin{table}[h]
\centering
\caption{Hyper-parameters for the single pendulum system. }
\label{tab:hyper_sp}
\begin{tabular}{@{}lcccc@{}}
\toprule
Hyper-parameters & \multicolumn{3}{c}{Model} \\
\cmidrule(r){2-4}
                  & \updatedTextRebutalTwo{MBD-NODE}          &LSTM & FCNN \\ \midrule
No. of layers    & \updatedText{2}          & \updatedText{2}                             & \updatedText{2}                        \\
\updatedText{No. of nodes per hidden layer}  & \updatedText{256} & \updatedText{256}& \updatedText{256}     \\
Max. epochs                  & 400       & 400                     & 600                  \\
Initial learning rate &  6e-4 &5e-4 & 5e-4 \\
Learning rate decay & 0.98  &0.98    &  0.98       \\
Activation function & Tanh &  Sigmoid,Tanh &Tanh\\
Loss function                  & MSE   & MSE& MSE                    \\
Optimizer                       & Adam      & Adam  & Adam                                \\  \bottomrule
\end{tabular}
\end{table}

\updatedText{
Figures \ref{fig:sp_xvt} and \ref{fig:sp_phase} present the dynamics response and the phase space of the single pendulum system during the ID generalization and OOD generalization. 
The \updatedTextRebutalTwo{MBD-NODE} outperforms other models with an MSE $\epsilon$= 2.0e-3. Although LSTM has a small MSE $\epsilon$= 3.4e-3, it tends to replicate some of the historical patterns and fails to capture the damping effect for OOD generalization.
The FCNN model has a larger MSE $\epsilon$= 8.0e-1, associated with the lackluster OOD generalization ability of the FCNN model.}

\begin{figure}[htbp]
    \centering
    \includegraphics[width=12cm]{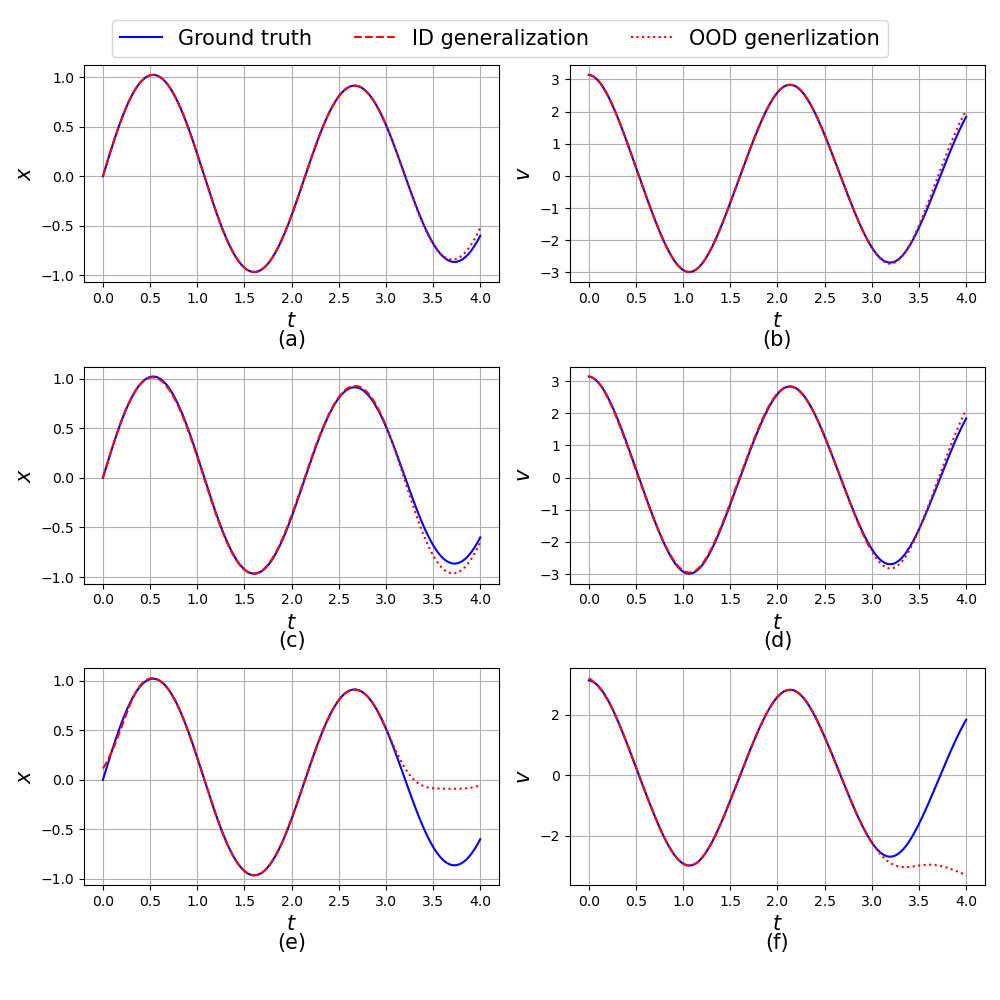}
    \caption{Comparison of $t$ vs $x$ (Left Column) and $t$ vs $v$ (Right Column) for the different models (rows) for the single pendulum system. Notice the dashed lines represent performance on the training data set $t \in [0,3]$, after which the dotted lines represent performance on the testing data set. (a), (b) are for the \updatedTextRebutalTwo{MBD-NODE} with MSE $\epsilon=$ 2.0e-3; (c), (d) are for the LSTM with MSE $\epsilon=$ 3.4e-3; (e), (f) are for the FCNN with MSE $\epsilon=$ 8.0e-1.}
    \label{fig:sp_xvt}
\end{figure}
\begin{figure}[htbp]
    \centering
    \includegraphics[width=12cm]{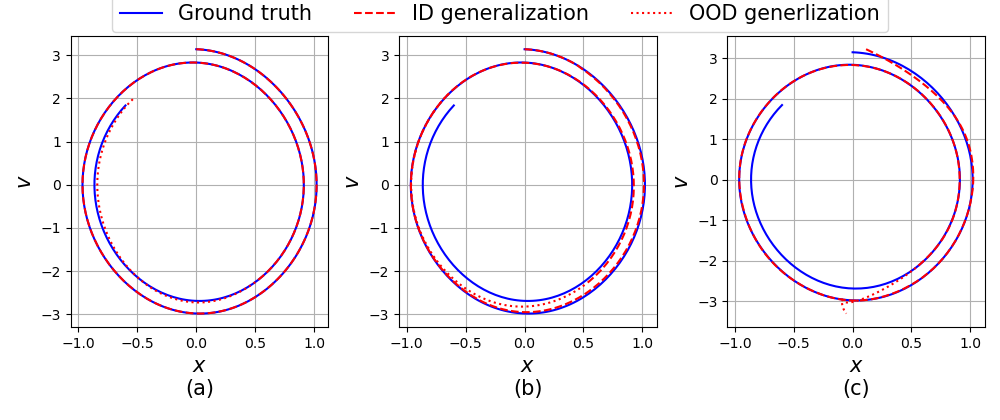}
    \caption{The phase space trajectories for the single pendulum system. Dashed lines represent performance on the training data and the dotted lines on the test data. (a) is for the \updatedTextRebutalTwo{MBD-NODE}; (b) is for the LSTM; (c) is for the FCNN.}
    \label{fig:sp_phase}
\end{figure}
\updatedText{
Beyond the first setting, we test \updatedTextRebutalTwo{MBD-NODE}'s ability to generalize under varying initial conditions and external forces. Importantly, we use the \updatedTextRebutalTwo{MBD-NODE} trained in the first setting directly without adding new training data – a significant challenge for OOD generalization. FCNNs and LSTMs are not suitable for handling time-varying external forces. They can only work with different parameters that do not change with respect to time.  For FCNNs, accommodating changes in initial conditions would require a larger model, additional data, and retraining.  Therefore, we only test \updatedTextRebutalTwo{MBD-NODE} in this setting. Additionally, \updatedTextRebutalTwo{MBD-NODE}'s nature allows us to directly calculate acceleration from external forces and incorporate it, simplifying integration with gravity as shown in Fig. \ref{fig: MNODE_un}.
}

\updatedText{
Figure \ref{fig: sp_initialization} presents the dynamics response of the single pendulum system with four different unseen initial conditions given in four quadrants. Because the \updatedTextRebutalTwo{MBD-NODE} learns the dynamics related to the range of the phase space covered in the training set and does so independently of the initial condition, \updatedTextRebutalTwo{MBD-NODE} yields a reasonable prediction for any of the four different initial conditions.}
\begin{figure}
    \centering
    \includegraphics[width=12cm]{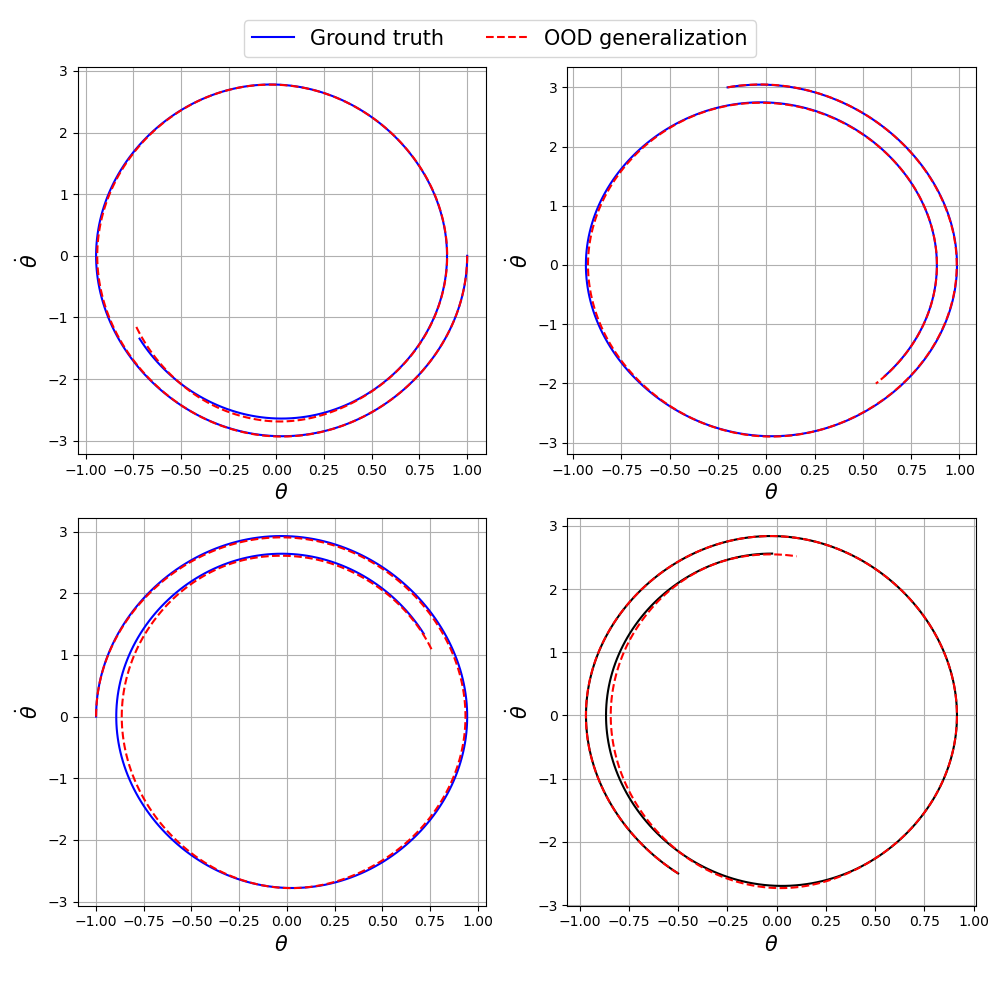}
    \caption{The prediction trajectory $\theta$ vs $\omega$ plot for the single pendulum by \updatedTextRebutalTwo{MBD-NODE} with different initialization: (a): $(\theta(0),\omega(0))=(1,0)$; (b): $(\theta(0),\omega(0))=(-0.2,,3)$; (c): $(\theta(0),\omega(0))=(-1,0)$; (d): $(\theta(0),\omega(0))=(-0.5,-2.5)$.} 
    \label{fig: sp_initialization}
\end{figure}
\updatedText{
Figure \ref{fig: sp_force} presents the dynamics response of the single pendulum system with random external force. Here, we sample the external force from the normal distribution $F \sim \mathcal{N}(0,25)$ and apply it to the single pendulum at every time step. We predict 300 time steps for the single pendulum with random excitation. Because the force will push the pendulum to unseen state space, this is a good test to probe the \updatedTextRebutalTwo{MBD-NODE}'s OOD generalization ability.
\updatedTextRebutalTwo{MBD-NODE} can continue to give an accurate prediction for the single pendulum system under random external force.}
\begin{figure}
    \centering
    \includegraphics[width=12cm]{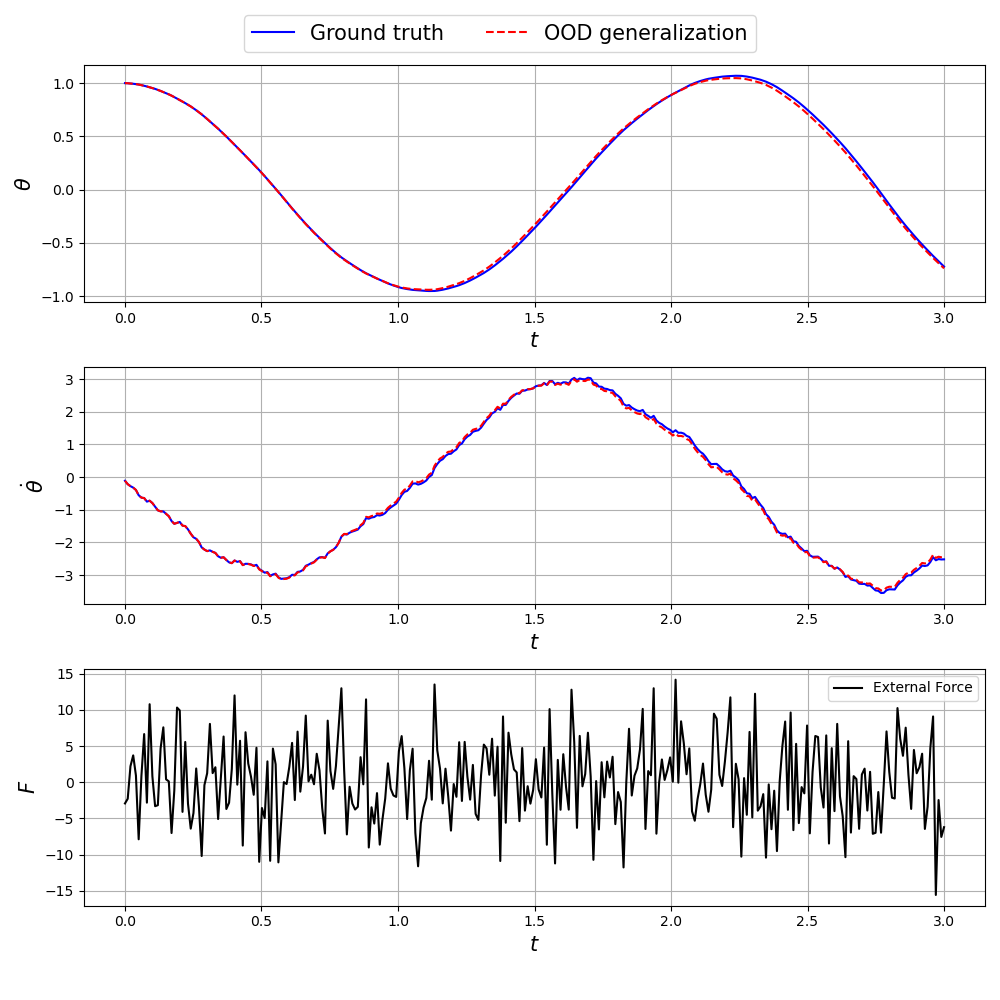}
    \caption{The dynamics response plot for $t$ vs $\theta$, $t$ vs $\omega$ plot and the value of random force applied to the single pendulum.}
    \label{fig: sp_force}
\end{figure}

\updatedText{\subsection{Double Pendulum}}
To gauge the performance of our model on chaotic systems, we study the double pendulum system (see Fig. \ref{fig:double_pendulum}) as a numerical example. This pendulum system has two masses \( m_1 \) and \( m_2 \), lengths \( l_1 \) and \( l_2 \), and two angles \( q_1=\theta_1 \) and \(q_2= \theta_2 \). The generalized momenta corresponding to these angles are \( p_{\theta_1} \) and \( p_{\theta_2} \), which needs to be calculated by using the Lagrangian. The Hamiltonian \( H \) for this system is given by:
\updatedTextRebutalTwo{
\begin{equation}
H(q_1,q_2,p_{\theta_1},p_{\theta_2} ) = T(q_1,q_2,p_{\theta_1},p_{\theta_2}) + V(q_1,q_2),
\end{equation}
where \( T \) is the kinetic energy and \( V \) is the potential energy.
}
For the double pendulum system, the kinetic energy \( T \) and potential energy \( U \) are given by:
\updatedText{
\begin{align}
T &= \frac{1}{2} m_1 l_1^2 \dot{\theta}_1^2 + \frac{1}{2} m_2 \left( l_1^2 \dot{\theta}_1^2 + l_2^2 \dot{\theta}_2^2 + 2 l_1 l_2 \dot{\theta}_1 \dot{\theta}_2 \cos(\theta_1 - \theta_2) \right), \\
U &= -m_1 g l_1 \cos(\theta_1) - m_2 g \left( l_1 \cos(\theta_1) + l_2 \cos(\theta_2) \right).
\end{align}}

The Hamiltonian \( H \) can be expressed in terms of\updatedTextRebutalTwo{ \( q_1, q_2, p_{\theta_1}, \) and \( p_{\theta_2} \)} as:
\updatedTextRebutalTwo{
\begin{equation}
\begin{aligned}
H(q_1, q_2, p_{\theta_1}, p_{\theta_2}) = & \frac{p_{\theta_1}^2}{2 m_1 l_1^2} + \frac{p_{\theta_2}^2}{2 m_2 l_2^2} + m_2 l_1 l_2 \cos(q_1 - q_2) p_{\theta_1} p_{\theta_2} \\
& - m_1 g l_1 \cos(q_1) - m_2 g \left( l_1 \cos(q_1) + l_2 \cos(q_2) \right)
\end{aligned}
\end{equation}}

The gradients of the Hamiltonian, \( \nabla_q H \) and \( \nabla_p H \), can be used to derive the Hamilton's equations of motion:
\updatedTextRebutalTwo{
\begin{align}
\dot{q}_i &= \frac{\partial H}{\partial p_{\theta_i}}, \\
\dot{p_{\theta_i}} &= -\frac{\partial H}{\partial q_i}.
\end{align}
}

The specific Hamilton's equations of motion for the double pendulum system are:
\updatedText{
\begin{align}
\dot{\theta}_1 &= \frac{l_2 p_{\theta_1} - l_1 p_{\theta_2} \cos(\theta_1 - \theta_2)}{l_1^2 l_2 \left[ m_1 + m_2 \sin^2(\theta_1 - \theta_2) \right]} \\
\dot{\theta}_2 &= \frac{-m_2 l_2 p_{\theta_1} \cos(\theta_1 - \theta_2) + (m_1 + m_2) l_1 p_{\theta_2}}{m_2 l_1 l_2^2 \left[ m_1 + m_2 \sin^2(\theta_1 - \theta_2) \right]} \\
\dot{p}_{\theta_1} &= -(m_1 + m_2)g l_1 \sin \theta_1 - h_1 + h_2 \sin \left[ 2(\theta_1 - \theta_2) \right] \\
\dot{p}_{\theta_2} &= -m_2 g l_2 \sin \theta_2 + h_1 - h_2 \sin \left[ 2(\theta_1 - \theta_2) \right]
\end{align}
}

where:
\updatedText{
\begin{align}
h_2 &= \frac{m_2 l_2^2 p_{\theta_1}^2 + (m_1 + m_2) l_1^2 p_{\theta_2}^2 - 2 m_2 l_1 l_2 p_{\theta_1} p_{\theta_2} \cos(\theta_1 - \theta_2)}{2 l_1^2 l_2^2 \left[ m_1 + m_2 \sin^2(\theta_1 - \theta_2) \right]^2} \\
h_1 &= \frac{p_{\theta_1} p_{\theta_2} \sin(\theta_1 - \theta_2)}{l_1 l_2 \left[ m_1 + m_2 \sin^2(\theta_1 - \theta_2) \right]}
\end{align}
}

\begin{figure}
    \centering
    \begin{tikzpicture}[>=latex]
        \def\rodOneLength{3.0}
        \def\rodTwoLength{2.5}
        \def\massOneRadius{0.2}
        \def\massTwoRadius{0.2}
        \def\pivotRadius{0.05}
        \def\angleOne{45} 
        \def\angleTwo{30} 
        \def\axissize{2.5} 

        \draw[->,thick,dashed] (-1,0) -- (\axissize,0) node[right] {$\displaystyle \mathbf{x}$};
        \draw[->,thick,dashed] (0,0) -- (0,-\rodOneLength-\rodTwoLength-\massTwoRadius) node[below] {$\displaystyle \mathbf{y}$};

        \fill (0,0) circle (\pivotRadius) node[above] {};

        \path (0,0) -- ++(-\angleOne:\rodOneLength) coordinate (m1);

        \draw[thick] (0,0) -- (m1) node[midway,left] {$\displaystyle \mathbf{l_1}$};

        \shade[ball color=black] (m1) circle (\massOneRadius) node[below left] {$\displaystyle \mathbf{m_1}$};

        \path (m1) -- ++(-\angleOne-\angleTwo:\rodTwoLength) coordinate (m2);

        \draw[thick] (m1) -- (m2) node[midway,below right] {$\displaystyle \mathbf{l_2}$};

        \shade[ball color=black] (m2) circle (\massTwoRadius) node[below left] {$\displaystyle \mathbf{m_2}$};
    
        \draw ([shift=(270:\rodOneLength/4)]0,0) arc (270:270+\angleOne:\rodOneLength/4);
        \node at (270+\angleOne*0.5:\rodOneLength/3) {$\displaystyle \mathbf{\theta_1}$};

        \draw ([shift=(270:\rodOneLength/4)]m1) arc (270:270+\angleTwo/2:\rodOneLength/4);
        \node at ([shift=(-\angleOne-\angleTwo*0.5:\rodOneLength/3)]m1) {$\displaystyle \mathbf{\theta_2}$};

        \draw[dashed] (0,0) -- (0,-\rodOneLength-\rodTwoLength-\massTwoRadius);
        \draw[dashed] (m1) -- ($(m1)+(0,-1.5*\rodTwoLength)$);
    \end{tikzpicture}
    \caption{Double pendulum}
    \label{fig:double_pendulum}
\end{figure}
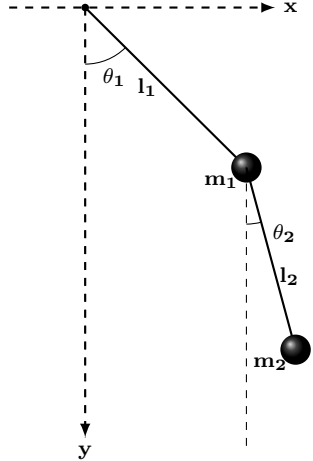

The double pendulum system is defined as follows: rod lengths, \(L_1 = L_2 = 1\) m; concentrated masses, \(m_1 = m_2 = 1\) kg; gravitational acceleration, \(g = 9.81\) m/s\(^2\); initial angular displacement of the first mass, \(\theta_1(0) = \frac{3\pi}{7}\); initial angular velocity of the first mass, \(\dot{\theta}_1(0) = 0\);
initial angular displacement of the second mass \(\theta_2(0) = \frac{3\pi}{4}\); initial angular velocity of the second mass \(\dot{\theta}_2(0) = 0\). We set the time step as 0.01s for both training and testing. The training dataset has a trajectory numerically computed via RK4 for 300 time steps. The models are tested by extrapolating for 100 more time steps. The hyperparameters used for the models are summarized in Table \ref{tab:hyper_dp}.

\begin{table}[h]
\centering
\caption{Hyper-parameters for the double pendulum system}
\label{tab:hyper_dp}
\begin{tabular}{@{}lcccc@{}}
\toprule
Hyper-parameters & \multicolumn{3}{c}{Model} \\
\cmidrule(r){2-4}
                  & \updatedTextRebutalTwo{MBD-NODE}          &LSTM & FCNN \\ \midrule
No. of hidden layers     & \updatedText{2}             & \updatedText{2}                         & \updatedText{2}                      \\
\updatedText{No. of nodes per hidden layer}  & \updatedText{256} & \updatedText{256} & \updatedText{256}  \\

Max. epochs                  & 450       & 400                     & 600                  \\
Initial learning rate &  1e-3 &5e-4 & 5e-4 \\
Learning rate decay & 0.98  &0.98    &  0.99       \\ 
Activation function & Tanh &  Sigmoid,Tanh &Tanh\\
Loss function                  & MSE & MSE& MSE                   \\
Optimizer                        & Adam     & Adam& Adam                                  \\  \bottomrule
\end{tabular}
\end{table}

Figure~\ref{fig:dp_xvt} shows the dynamic response of the different methods for a double pendulum. We can observe that all three models can give good predictions in the range of the training set. 
In the extrapolation range, the three models gradually diverge. 
\updatedText{ The challenge for integrator-based methods like \updatedTextRebutalTwo{MBD-NODE} is particularly pronounced due to the inherently chaotic nature of the double pendulum system, which tends to amplify integration errors rapidly, leading to significant discrepancies.  
For a discussion about the limitations of numerical integration methods like the Runge-Kutta and integration-based neural networks like Physics-Informed Neural Networks (PINN), the reader is referred to \cite{PINN_cheats}. For the double pendulum problem, these two approaches give large divergence for small initial perturbation. Despite this, the \updatedTextRebutalTwo{MBD-NODE} outperforms the two other models with an MSE of $\epsilon=$2.0e-1.}


The phase space trajectories obtained by the three models are shown in Fig.~\ref{fig:dp_phase}. We can observe that the \updatedTextRebutalTwo{MBD-NODE} model overall outperforms the other two models in the testing data regime. Although there are noticeable differences between the prediction and ground truth for the \updatedTextRebutalTwo{MBD-NODE} model, it's still trying to capture the patterns of ground truth in the testing regime, especially for the second mass. On the contrary, the LSTM model tends to replicate the historical trajectories as shown in Fig. \ref{fig:dp_phase}(c) and (d). For example, the FCNN fails to demonstrate predictive attributes outside of the training regime.

\begin{figure}[htbp]
    \centering
    \includegraphics[width=11cm]{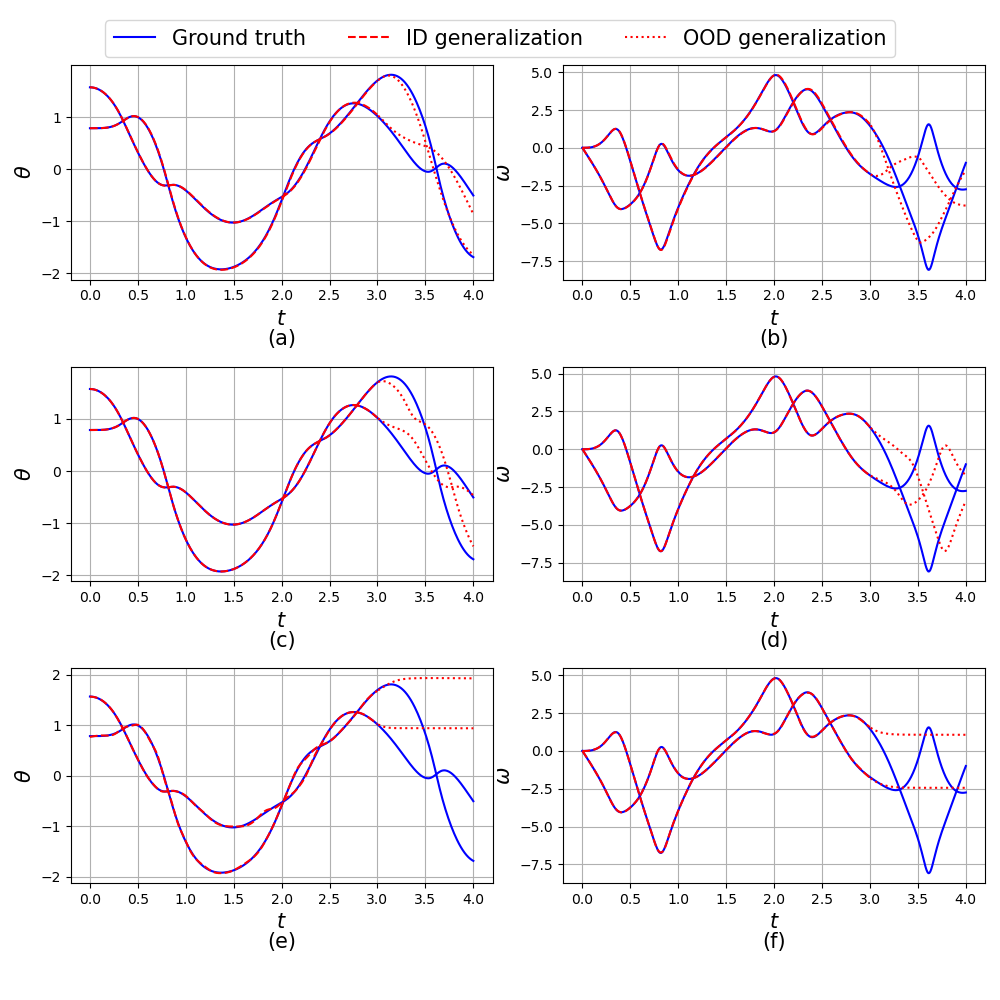}
    \caption{Comparison of $t$ vs $\theta$ (Left Column) and $t$ vs $\omega$ (Right Column) for the different models (rows) for the double pendulum system. Notice the dashed lines represent performance on the training data set $t \in [0,3]$ after which the dotted lines represent performance on the testing data set. (a), (b) are for the \updatedTextRebutalTwo{MBD-NODE} with MSE $\epsilon=$2.0e-1; (c), (d) are for the LSTM with MSE $\epsilon=$6.4e-1; (e), (f) are for the FCNN with MSE $\epsilon=$ 2.2e+0.}
    \label{fig:dp_xvt}
\end{figure}

\begin{figure}[htbp]
    \centering
    \includegraphics[width=11cm]{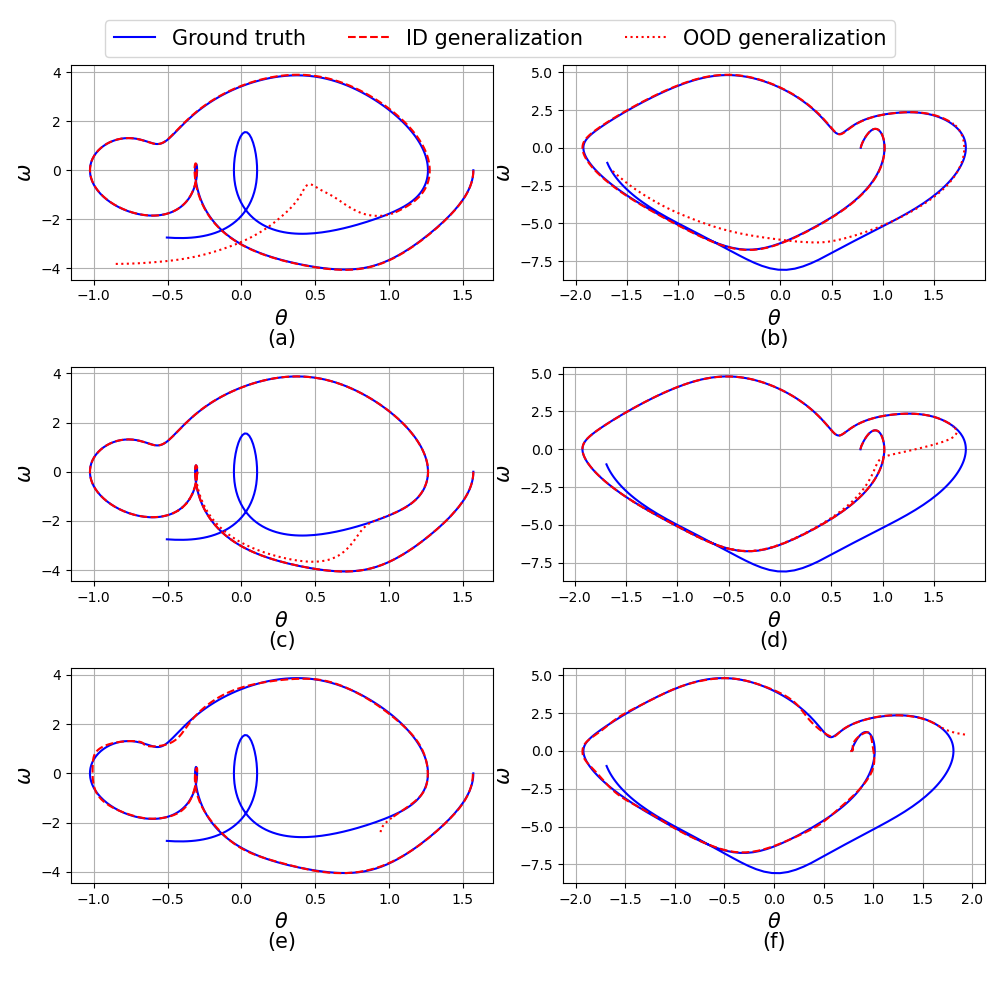}
    \caption{The phase space trajectories for the double pendulum system. The left and right columns correspond to the first and second masses, respectively. Dashed lines represent performance on the training data and the dotted lines on the test data. (a), (b) are for the \updatedTextRebutalTwo{MBD-NODE}; (c), (d) are for the LSTM; (e), (f) are for the FCNN.}
    \label{fig:dp_phase}
\end{figure}

\updatedText{\subsection{Cart-pole System}}
\updatedText{
In this section, we consider the cart-pole system, which is a classical benchmark problem in control theory. 
As shown in Fig.~\ref{fig:cart-pole}, the system consists of a cart that can move horizontally along a frictionless track and a pendulum that is attached to the cart. 
The pendulum is free to rotate about its pivot point. 
The system's state is described by the position of the cart \(x\), the velocity of the cart \(v\), the angle of the pendulum \(\theta\), and the angular velocity of the pendulum \(\omega\). 
The equations of motion for the cart-pole system are given by the following second-order nonlinear ODEs:
}

\begin{equation}
\begin{aligned}
    ml^2\ddot{\theta} + ml\cos{\theta}\ddot{x} & -mgl\sin{\theta} = 0 \\
    ml\cos{\theta}\ddot{\theta} + (M + m)\ddot{x} &- ml\dot{\theta}^2\sin{\theta} = u,
\end{aligned}\label{eqn:cart-pole}
\end{equation}
\updatedText{where $M=1\ \unit{kg}$ is the mass of the cart and $m=1\ \unit{kg}$ is the mass of the pole, $l=0.5\ \unit{m}$ is the length of the pole, and $u$ is the external force horizontally applied to the cart with unit $N$. 
}
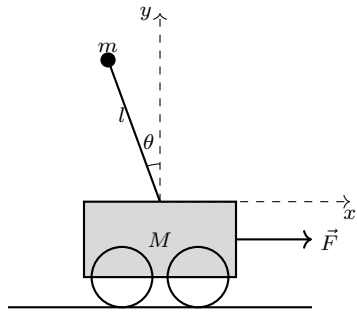
\begin{figure}
\centering
\begin{tikzpicture}
    \def\cartwidth{2.0}
    \def\cartheight{1.0}
    \def\pendulumlength{2.0}
    \def\pendulumwidth{0.1}
    \def\wheelradius{0.4}

    \draw[thick, fill=gray!30] (-\cartwidth/2,0) rectangle (\cartwidth/2,-\cartheight);
    
    \draw[thick] (-\cartwidth/4,-\cartheight) circle (\wheelradius);
    \draw[thick] (\cartwidth/4,-\cartheight) circle (\wheelradius);
    
    \draw[thick] (0,0) -- ++(110:\pendulumlength) node[midway, above left] {$l$};
    \fill (110:\pendulumlength) circle (0.1) node[above] {$m$};

    \draw (0,0.5) arc (90:110:0.5);
    \node at (-0.15,0.8) {$\theta$};
    
    \draw[->,thick] (\cartwidth/2,-\cartheight/2) -- +(1,0) node[right] {$\vec{F}$};
    
    \draw[->,dashed] (0,0) -- ++(0,2.5) node[left] {$y$};
    \draw[->,dashed] (0, 0) -- (2.5, 0) node[below] {$x$};
    
    \node at (0,-\cartheight/2) {$M$};

    \draw[thick] (-\cartwidth, -\cartheight-\wheelradius) -- (\cartwidth, -\cartheight-\wheelradius);
\end{tikzpicture}
\caption{The cart-pole system.}
\label{fig:cart-pole}
\end{figure}
\updatedText{
We first consider the case in which the cart-pole system is set to an initial position, and then we let the system evolve without any external force being applied. 
The initial conditions are set as follows:}
\updatedText{
\begin{equation}
    x(0)= 1,\quad v(0)= 0,\quad \theta(0)= \frac{\pi}{6},\quad \omega(0) = 0.
\label{eqn:cart-pole_init}
\end{equation}}
\updatedText{
The system is simulated using the midpoint method with a time step of 0.005s. 
We generate the training data by simulating the system for 400 time steps and the testing data by simulating the system for 100 time steps.
As shown in Figs.~\ref{fig:cart-pole_x_v_t} and \ref{fig:cart-pole_phrase}, the \updatedTextRebutalTwo{MBD-NODE} can accurately predict the system dynamics with $\sigma=6.0e-5$. 
Because this case is a periodic system that time series data can fully describe, the LSTM model can also provide accurate predictions with $\sigma=3.0e-4$. 
However, the FCNN model still gives lackluster OOD generalization performance with $\sigma=4.7e-2$. The hyperparameters for each model are summarized in Table \ref{tab:hyper_cp}.}
\begin{table}[h]
    \centering
    \caption{Hyper-parameters for the cart-pole system}
    \label{tab:hyper_cp}
    \begin{tabular}{@{}lcccc@{}}
    \toprule
    \updatedText{Hyper-parameters} & \multicolumn{3}{c}{\updatedText{Model}} \\
    \cmidrule(r){2-4}
                      & \updatedText{\updatedTextRebutalTwo{MBD-NODE}}          &\updatedText{LSTM} & \updatedText{FCNN} \\ \midrule
                      \updatedText{No. of hidden layers}     &\updatedText{2}             & \updatedText{2}                             & \updatedText{2}                          \\
                      \updatedText{No. of nodes per hidden layer}  & \updatedText{256}   &  \updatedText{256} & \updatedText{256}  \\
    
                      \updatedText{Max. epochs}                  & \updatedText{450 }      & \updatedText{400 }                    &\updatedText{ 600}                  \\
                      \updatedText{Initial learning rate} & \updatedText{ 1e-3} &\updatedText{5e-4} & \updatedText{5e-4} \\
                      \updatedText{Learning rate decay} &\updatedText{ 0.98}  &\updatedText{0.98 }   &  \updatedText{0.99}       \\ 
                      \updatedText{Activation function} & \updatedText{Tanh} &  \updatedText{Sigmoid,Tanh} &\updatedText{Tanh}\\
                      \updatedText{Loss function}                  & \updatedText{MSE} & \updatedText{MSE}& \updatedText{MSE}                   \\
                      \updatedText{Optimizer}                        & \updatedText{Adam}     & \updatedText{Adam}& \updatedText{Adam}                                  \\  \bottomrule
    \end{tabular}
    \end{table}

\begin{figure}[h]
    \centering
    \includegraphics[width=12cm]{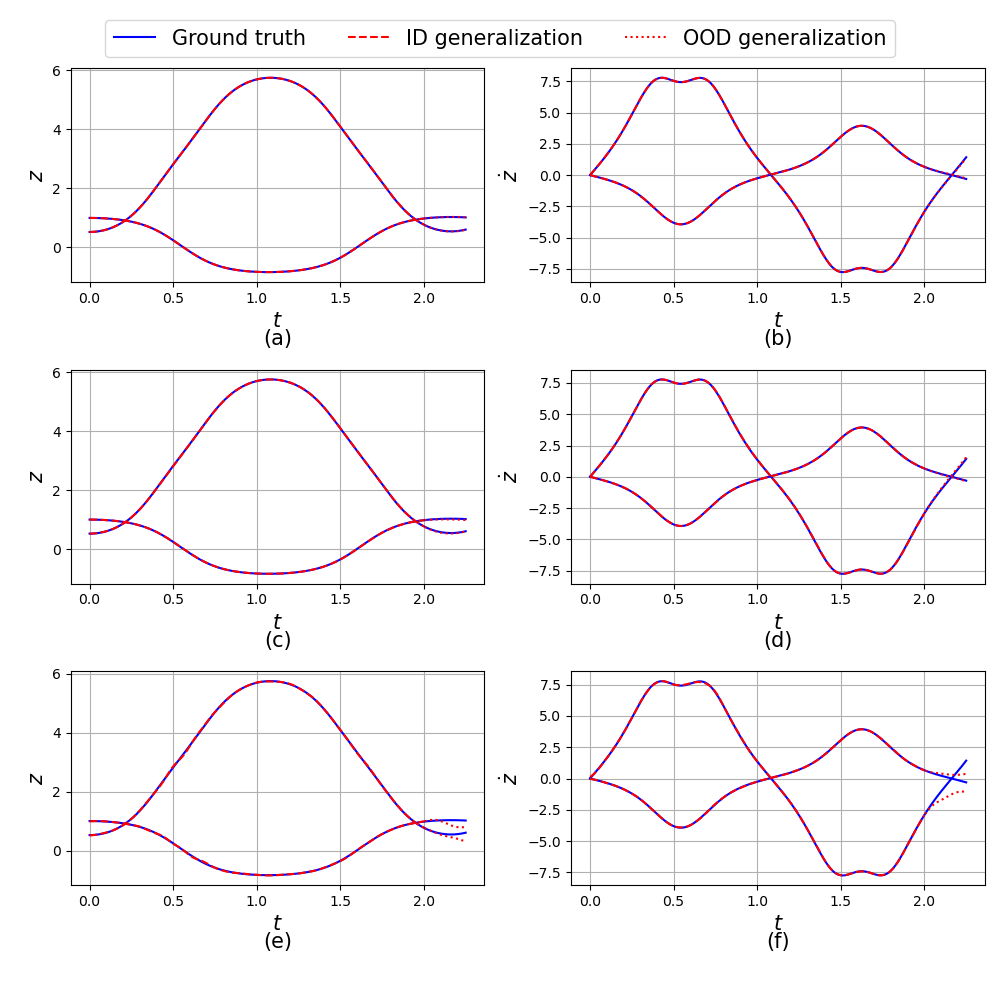}
    \caption{Comparison of $t$ vs $z$ (Left Colum) and $t$ vs $\dot z$ (Right Column) for the different models (rows) for the cart-pole system. The dashed lines represent the ID generation, and the dotted lines represent the OOD generalization. (a), (b) are for the \updatedTextRebutalTwo{MBD-NODE} with MSE $\sigma=6.0e-5$; (c), (d) are for the LSTM with $\sigma=3.2e-4$; (e), (f) are for the FCNN with $\sigma=4.7e-2$.}
    \label{fig:cart-pole_x_v_t}
\end{figure}

\begin{figure}[h]
    \centering
    \includegraphics[width=12cm]{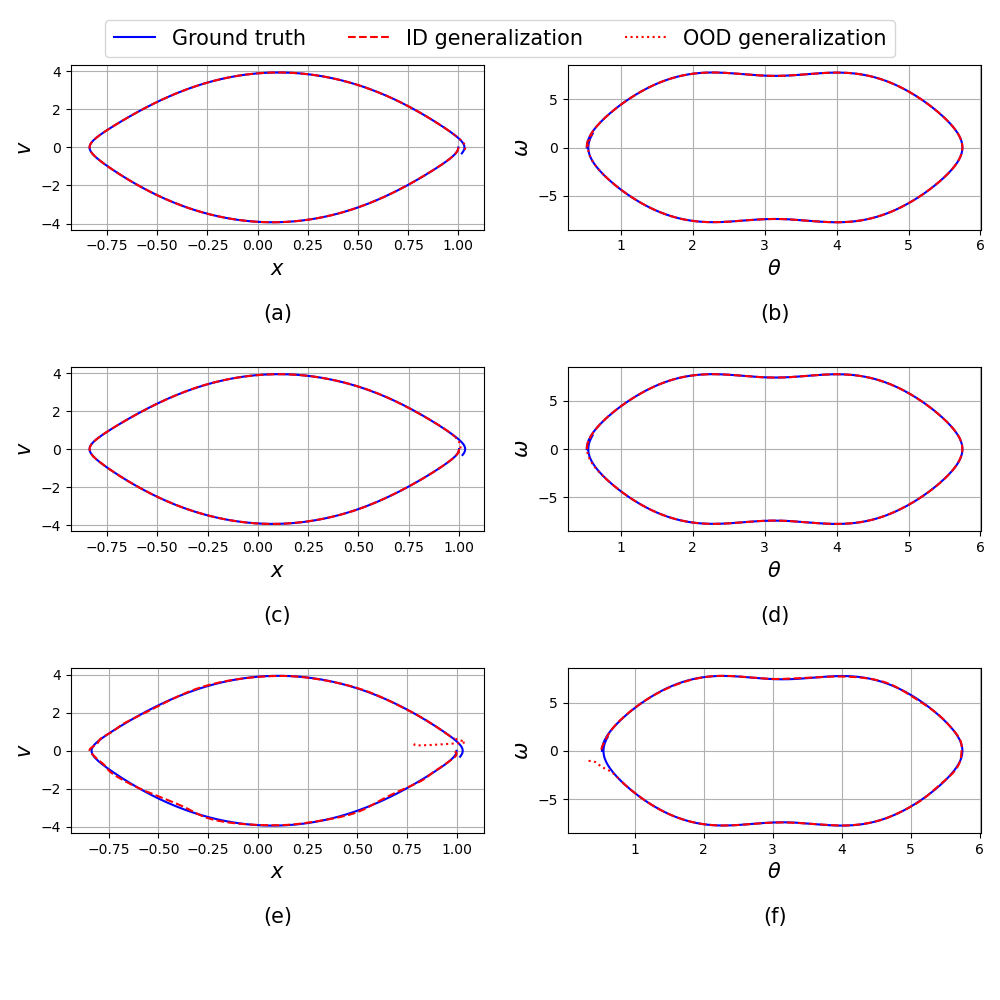}
    \caption{The phase space $x$ vs $v$ (Left Column) and $\theta$ vs $\dot \theta$ (Right Column) of the cart-pole system. The dashed line represents the ID generation, and the dotted line represents the ODD generalization. (a), (b) are for the \updatedTextRebutalTwo{MBD-NODE}; (c), (d) are for the LSTM; (e), (f) are for the FCNN. }
    \label{fig:cart-pole_phrase}
\end{figure}
\updatedText{
Furthermore, we consider the case that the cart-pole system is set to the initial position Eq.~\eqref{eqn:cart-pole_init}, and then we apply the external force $u$ to the cart to balance the pole and keep the cart-pole system at the origin point.
In general control theory, model predictive control (MPC) is a popular method for solving this kind of control problem by linearizing the nonlinear system dynamics and solving a quadratic convex optimization problem over a finite time horizon at each time step.
Specifically, for a linearized system dynamics $\dot z=Az+Bu$, the optimization problem can be formulated as a convex optimization problem as follows \cite{kouvaritakis2016MPC}:}

    \begin{align}
        \min_{u} & \sum_{k=0}^{N-1}z_k^TQz_k+u_k^TRu_k\\
        \text{s.t.} & z_{k+1}=Az_k+Bu_k, \quad k=0,1,\cdots,N-1 \\
        & z_k\in Z, \quad u_k\in U, \quad k=0,1,\cdots,N-1,
    \end{align}
    \updatedText{
where $z_k=(\theta_k,x_k,\omega_k,v_k)$ is the state of the system at time step $k$, $N=50$ is the time horizon for optimization,
 $u_k$ is the control input at time step $k$, $Q$ and $R$ are the weighting matrices, which are set to the identity matrix, and $Z$ and $U$ are the constraints for the state and control input, respectively.}

 \updatedText{
For the cart-pole system, the matrix $A$ and $B$ can be easily derived from the system dynamics Eq.~\ref{eqn:cart-pole} by the first-order Taylor series approximation, which are:}

\begin{equation}
    A=\begin{bmatrix}
        0 & 0 & 1 & 0\\
        0 & 0 & 0 & 1\\
        \frac{g(m+M)}{Ml} & 0 & 0 & 0\\
        -\frac{mg}{M} & 0 & 0 & 0
    \end{bmatrix}, \quad B=\begin{bmatrix}
        0\\
        0\\
        -\frac{1}{Ml}\\
        \frac{1}{M}\\
    \end{bmatrix}.
\end{equation}
\updatedText{
As a high-accuracy and differentiable model, the \updatedTextRebutalTwo{MBD-NODE} can be used to directly linearize the system dynamics by calculating the Jacobian matrix of the well-trained \updatedTextRebutalTwo{MBD-NODE}.
In this case, \updatedTextRebutalTwo{MBD-NODE} captures the system dynamics by learning the map $f(\theta_k,x_k,\omega_k,v_k,u)$ to the angular acceleration $\ddot \theta_k$ and the acceleration $\ddot x_k$.
In practice, the Jacobian matrix can be calculated by automatic differentiation, which is used to replace the matrix $A$ and $B$ in the MPC optimization problem.
To get the well-trained \updatedTextRebutalTwo{MBD-NODE}, we train the model with $10^5$ uniformly sampling data points in the range of $(\theta_k,x_k,\omega_k,v_k,u)\in [0,2\pi]\times[-1.5,1.5]\times[-8,8]\times[-4,4]\times[-10,30]$ for the state space and the control input.
We limit our analysis to the \updatedTextRebutalTwo{MBD-NODE} model as FCNN and LSTM models cannot work with time-evolving external input.}

\updatedText{
Figure \ref{fig:cart-pole_con_loss} shows the trajectories and the obtained control input for the MPC methods and the \updatedTextRebutalTwo{MBD-NODE}-based MPC method. We can see that the \updatedTextRebutalTwo{MBD-NODE}-based MPC can provide high-accuracy control input and trajectory as the analytic equation of motion-based MPC, which also shows \updatedTextRebutalTwo{MBD-NODE}'s strong ability to capture the system dynamics.}
\begin{figure}[h]
    \centering
    \includegraphics[width=12cm]{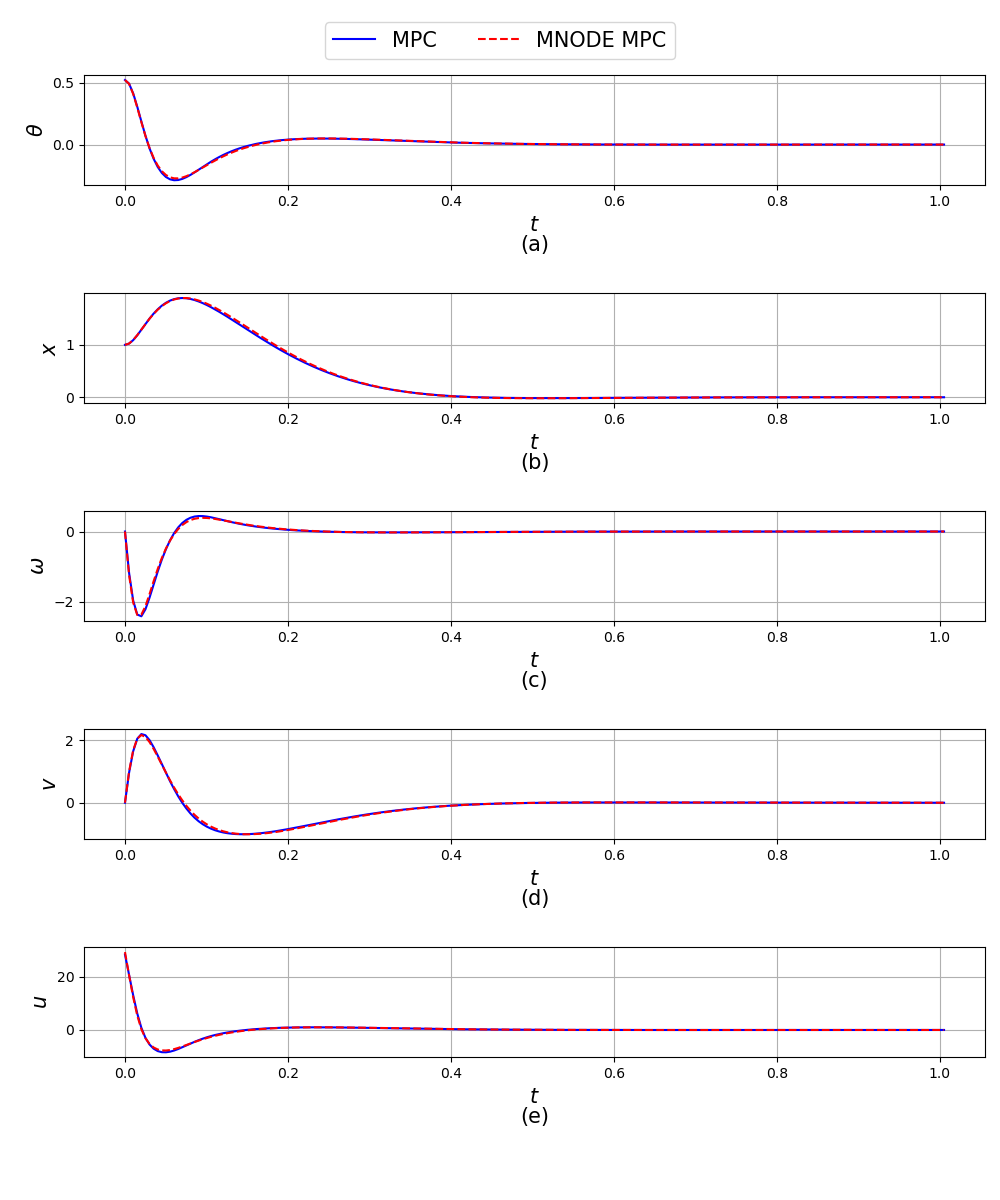}
    \caption{The trajectory and control input for the MPC methods and the \updatedTextRebutalTwo{MBD-NODE}-based MPC method. (a): the $\theta$ vs $t$; (b): the $x$ vs $t$; (c): the $\omega$ vs $t$. (d): the $v$ vs $t$; (e): the $u$ vs $t$.}
    \label{fig:cart-pole_con_loss}
\end{figure}

\updatedText{\subsection{Slider-Crank Mechanism}}
\label{sec:slider_crank}

\updatedText{
We assessed \updatedTextRebutalTwo{MBD-NODE}'s capacity for long-term, high-accuracy predictions using the slider-crank mechanism (Fig. \ref{fig:slider_crank}). The test involved generating predictions for up to 10,000 time steps (100s), while encompassing generalization to arbitrary external forces and torques applied to both the slider and crank.
We did not include LSTM and FCNN models in the comparison as their inherent structure does not readily accommodate the representation of system dynamics with variable external forces and torques.
Additionally, LSTM and FCNN models face challenges in long-term prediction. Their training data requirements and computational costs scale linearly with the time horizon, whereas \updatedTextRebutalTwo{MBD-NODE}'s performance depends on the phase space and external inputs, not directly on the time horizon.
Previous FCNN- and LSTM-based approaches in related work \cite{HAN_DNN2021113480,Choi_DDSdd57a6f7c2064450bc1f79231ef67414,YE_MBSNET2021107716,efficient_PCA} typically demonstrate short-term prediction capabilities, limited to durations of several seconds or hundreds of time steps.}

\begin{figure}[htbp]
    \centering
    \begin{tikzpicture}[x=0.75pt,y=0.75pt,yscale=-0.75,xscale=0.75]
    
    \draw  [dash pattern={on 4.5pt off 4.5pt}]  (200.5,83) -- (200.5,212) ;
    \draw [shift={(200.5,80)}, rotate = 90] [fill={rgb, 255:red, 0; green, 0; blue, 0 }  ][line width=0.08]  [draw opacity=0] (8.93,-4.29) -- (0,0) -- (8.93,4.29) -- cycle    ;
    \draw  [dash pattern={on 4.5pt off 4.5pt}]  (569.5,211.5) -- (201,211.5) ;
    \draw [shift={(572.5,211.5)}, rotate = 180] [fill={rgb, 255:red, 0; green, 0; blue, 0 }  ][line width=0.08]  [draw opacity=0] (8.93,-4.29) -- (0,0) -- (8.93,4.29) -- cycle    ;
    \draw    (255.5,137) -- (195.5,208) ;
    \draw    (265.5,146) -- (206.5,216) ;
    \draw  [draw opacity=0] (206.75,215.09) .. controls (206.46,215.83) and (206.03,216.53) .. (205.48,217.15) .. controls (202.92,220.02) and (198.53,220.28) .. (195.66,217.73) .. controls (192.78,215.18) and (192.52,210.78) .. (195.08,207.91) .. controls (195.63,207.28) and (196.28,206.78) .. (196.97,206.41) -- (200.28,212.53) -- cycle ; \draw   (206.75,215.09) .. controls (206.46,215.83) and (206.03,216.53) .. (205.48,217.15) .. controls (202.92,220.02) and (198.53,220.28) .. (195.66,217.73) .. controls (192.78,215.18) and (192.52,210.78) .. (195.08,207.91) .. controls (195.63,207.28) and (196.28,206.78) .. (196.97,206.41) ;  
    \draw  [draw opacity=0] (254.7,138.28) .. controls (254.94,137.53) and (255.32,136.81) .. (255.84,136.15) .. controls (258.21,133.13) and (262.58,132.6) .. (265.61,134.97) .. controls (268.63,137.34) and (269.16,141.71) .. (266.79,144.74) .. controls (266.27,145.39) and (265.66,145.93) .. (264.99,146.35) -- (261.31,140.44) -- cycle ; \draw   (254.7,138.28) .. controls (254.94,137.53) and (255.32,136.81) .. (255.84,136.15) .. controls (258.21,133.13) and (262.58,132.6) .. (265.61,134.97) .. controls (268.63,137.34) and (269.16,141.71) .. (266.79,144.74) .. controls (266.27,145.39) and (265.66,145.93) .. (264.99,146.35) ;  
    \draw   (196.25,211.5) .. controls (196.25,208.88) and (198.38,206.75) .. (201,206.75) .. controls (203.62,206.75) and (205.75,208.88) .. (205.75,211.5) .. controls (205.75,214.12) and (203.62,216.25) .. (201,216.25) .. controls (198.38,216.25) and (196.25,214.12) .. (196.25,211.5) -- cycle ;
    \draw   (256.56,140.44) .. controls (256.56,137.82) and (258.69,135.69) .. (261.31,135.69) .. controls (263.94,135.69) and (266.06,137.82) .. (266.06,140.44) .. controls (266.06,143.07) and (263.94,145.19) .. (261.31,145.19) .. controls (258.69,145.19) and (256.56,143.07) .. (256.56,140.44) -- cycle ;
    \draw    (264.99,146.35) -- (387.5,216) ;
    \draw    (266.5,135) -- (393.5,206) ;
    \draw  [draw opacity=0] (393.42,205.96) .. controls (393.42,205.96) and (393.42,205.96) .. (393.42,205.96) .. controls (393.42,205.96) and (393.42,205.96) .. (393.42,205.96) .. controls (396.24,207.59) and (397.21,211.2) .. (395.58,214.02) .. controls (393.94,216.84) and (390.33,217.81) .. (387.51,216.17) -- (390.47,211.06) -- cycle ; \draw   (393.42,205.96) .. controls (393.42,205.96) and (393.42,205.96) .. (393.42,205.96) .. controls (393.42,205.96) and (393.42,205.96) .. (393.42,205.96) .. controls (396.24,207.59) and (397.21,211.2) .. (395.58,214.02) .. controls (393.94,216.84) and (390.33,217.81) .. (387.51,216.17) ;  
    \draw   (385.98,211.06) .. controls (385.98,208.59) and (387.99,206.58) .. (390.47,206.58) .. controls (392.94,206.58) and (394.95,208.59) .. (394.95,211.06) .. controls (394.95,213.54) and (392.94,215.55) .. (390.47,215.55) .. controls (387.99,215.55) and (385.98,213.54) .. (385.98,211.06) -- cycle ;
    \draw    (350,181.5) -- (440.5,181) ;
    \draw    (339.5,240) -- (440,241) ;
  
    \draw    [thick](369.5,206) -- (369.08,239.5) ;
    \draw    [thick](370.08,181.5) -- (369.5,193) ;
    \draw    (410.08,181.5) -- (410.08,239.5) ;
    \draw  [draw opacity=0] (228.01,191.29) .. controls (230.18,192.25) and (232.15,196.69) .. (232.67,202.16) .. controls (233.02,205.84) and (232.63,209.19) .. (231.75,211.4) -- (228.05,202.59) -- cycle ; \draw   (228.01,191.29) .. controls (230.18,192.25) and (232.15,196.69) .. (232.67,202.16) .. controls (233.02,205.84) and (232.63,209.19) .. (231.75,211.4) ;  
    
\draw[->,thick] (201,211.5) ++(60:0.5cm) arc (420:120:0.5cm);
\node at (201,211.5) [above left=0.3cm and 0.3cm] {$\displaystyle \mathbf{T}$};

\draw[->,thick] (393.5,206) -- ++(1cm,0) node [above] {$\displaystyle \mathbf{F}$};

    \draw (551,198) node [anchor=north west][inner sep=0.75pt]   [align=left] {$\displaystyle \mathbf{x}$};
    \draw (205,91) node [anchor=north west][inner sep=0.75pt]   [align=left] {$\displaystyle \mathbf{y}$};

    \draw (232,192) node [anchor=north west][inner sep=0.75pt]   [align=left] {$\displaystyle \mathbf{\theta_1} $};
    \draw (209.12,175) node [anchor=north west][inner sep=0.75pt]  [rotate=-311.78] [align=left] {$\displaystyle \mathbf{r}$};
    \draw (320.9,148) node [anchor=north west][inner sep=0.75pt]  [rotate=-33.61] [align=left] {$\displaystyle \mathbf{l}$};
    \end{tikzpicture}
    \caption{The slider-crank system.}
    \label{fig:slider_crank}
\end{figure}
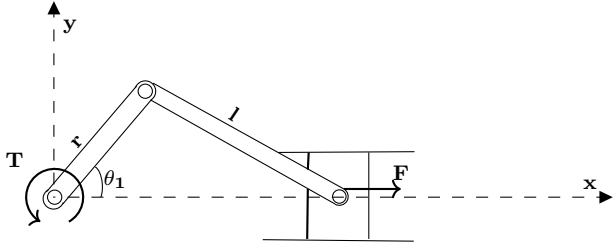

\updatedText{
We formulate the slider-crank mechanism as a three-body problem with hard constraints as follows:}

\begin{enumerate}
    \item  \updatedText{The crank is connected to ground with a revolute joint with mass $m_1 = 3 \, \unit{kg}$, moment of inertia $I_1 = 4 \, \text{kg} \cdot \text{m}^2$. The center of mass of the crank in the global reference frame is $(x_1, y_1,\theta_1)$, and the length of the crank is $l=2 \, \text{m}$.}
  
    \item  \updatedText{The rod is connected to the crank with a revolute joint with mass $m_2 = 6 \, \text{kg}$, moment of inertia $I_2 = 32 \, \text{kg} \cdot \text{m}^2$. The center of mass of the rod in the global reference frame is expressed as $(x_2, y_2,\theta_2)$, and the length of the connecting rod is $r=4 \, \text{m}$.}
    
    \item  \updatedText{The slider is connected to the rod with a revolute joint and constrained to move horizontally with mass $m_3 = 1 \, \text{kg}$, moment of inertia $I_3 = 1 \, \text{kg} \cdot \text{m}^2$. The center of mass of the slider in the global reference frame is $(x_3, y_3,\theta_3)$.}
\end{enumerate}
\updatedText{
The generalized coordinate $q=(x_1,y_1,\theta_1,x_2,y_2,\theta_2,x_3,y_3,\theta_3)$ are used to describe the system dynamics. Given at a point in time $q$, $\dot q$ and some values of the external force/torque $(F,T)$, we seek to produce the generalized acceleration $\ddot q$; i.e., we have a total of $9*2=18$ system states and $2$ external inputs to describe the system dynamics. Because the slider-crank mechanism is a one-DOF system, we take the minimum coordinates as $(\theta_1,\dot{\theta_1})$ with the external input $(F, T)$; these four variables fully determine the system dynamics. All other coordinates are treated as dependent coordinates. The detailed formulation is shown in the Appendix \ref{appsec:slider_crank}.}

\updatedText{For the training part, we uniformly sampled $10^7$ data points as $(\theta_1,\dot \theta_1, F, T) \in [0,2\pi]\updatedTextRebutalTwo{\times[-4,4]}\times[-10,10]\times[-10,10]$ providing the training data. The training used the hyperparameters shown in Table \ref{tab:hyper_sc}. In the testing part, we set the initial condition to be $({\theta_1},\dot {\theta_1})=(1,1)$, the simulation time step as 0.01s and the external force and torque $(F, T)\sim U[-10,10]\times U[-10,10]$ sampled from uniform distribution are applied to the system for each time step; note that there is no requirement for smoothness in $F$ and $T$, although if one is present that would only help. We run the prediction for 10000 steps (100s) to test the \updatedTextRebutalTwo{MBD-NODE}'s long-time prediction ability.}
\begin{figure}
    \centering
    \includegraphics[width=11cm]{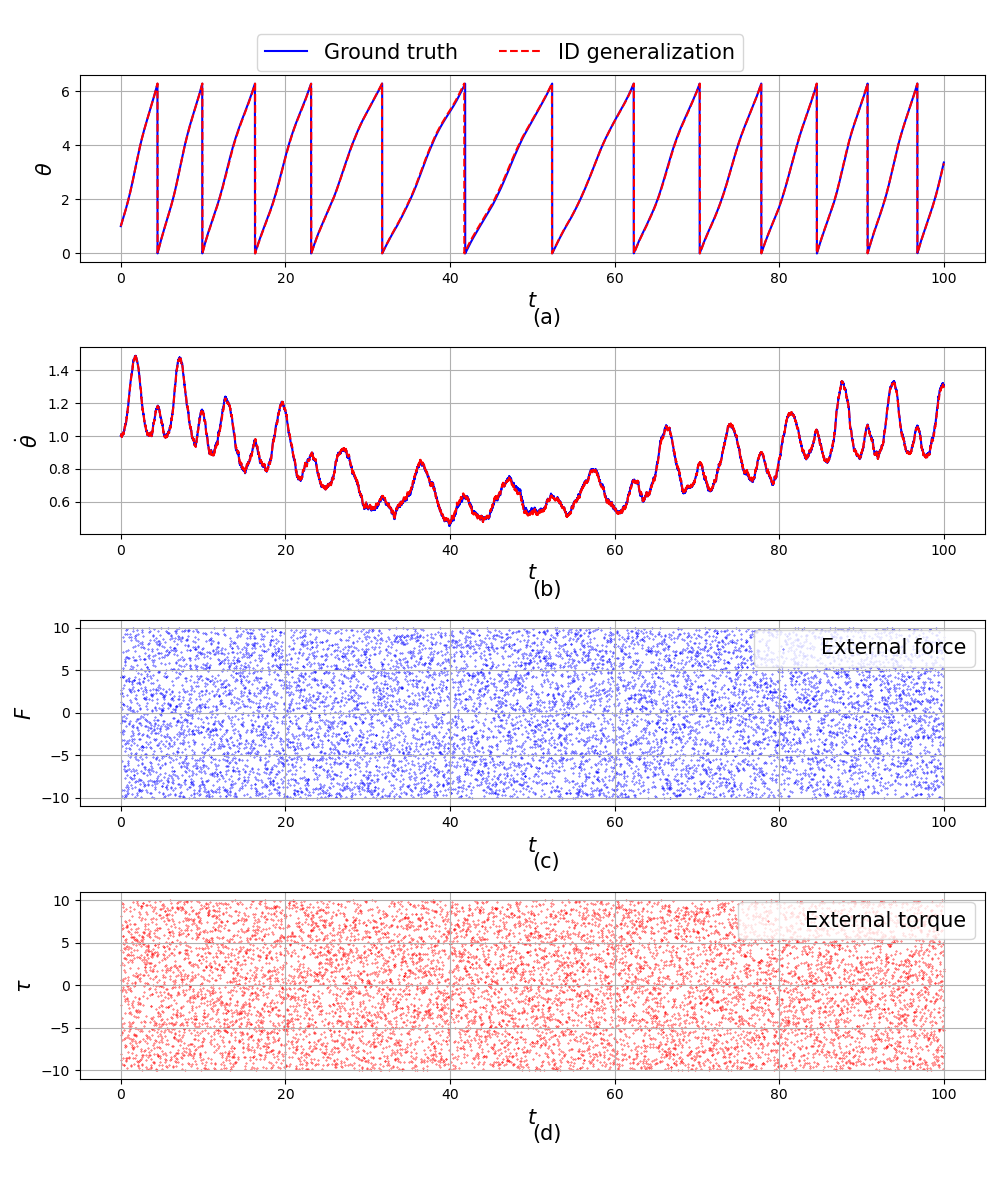}
    \caption{The dynamics response of the slider-crank mechanism under the external force and torque. (a): the $\theta_1$ vs $t$; (b): the $\dot \theta_1$ vs $t$; (c): the applied external force $F$ vs $t$; (d): the applied external torque $T$ vs $t$.}
    \label{fig:slider_crank_dyanmics}
\end{figure}

\begin{table}[h]
    \centering
    \caption{Hyper-parameters for the slider-crank mechanism.}
    \label{tab:hyper_sc}
    \begin{tabular}{@{}lcccc@{}}
    \toprule
    \updatedText{Hyper-parameters} & \multicolumn{3}{c}{ \updatedText{Model}} \\
    \cmidrule(r){2-4}
                      &  \updatedText{\updatedTextRebutalTwo{MBD-NODE}}          & \updatedText{LSTM} &  \updatedText{FCNN}\\ \midrule
                      \updatedText{No. of hidden layers}     &  \updatedText{2}          & \updatedText{-}                          &\updatedText{-}                \\
                      \updatedText{No. of nodes per hidden layer} & \updatedText{256}  & \updatedText{-}  & \updatedText{-}   \\
                      \updatedText{Max. epochs}                  &  \updatedText{500}       & \updatedText{-}                      & \updatedText{-}               \\
                      \updatedText{Initial learning rate} &  \updatedText{ 1e-3} &\updatedText{-}  & \updatedText{-} \\
                      \updatedText{Learning rate decay} &  \updatedText{0.98}  &\updatedText{-}    &  \updatedText{-}       \\ 
                      \updatedText{Activation function} &  \updatedText{Tanh} & \updatedText{-}  &\updatedText{-} \\
                      \updatedText{ Loss function}                  & \updatedText{ MSE} & \updatedText{-} & \updatedText{-}                   \\
                      \updatedText{Optimizer}                        &  \updatedText{Adam}     & \updatedText{-} & \updatedText{-}                                   \\  \bottomrule
    \end{tabular}
    \end{table}

\updatedText{Figure \ref{fig:slider_crank_dyanmics} shows the dynamics response of the minimal coordinates $(\theta_1,\dot \theta_1)$ under the external force and torque. \updatedTextRebutalTwo{MBD-NODE} accurately predicts the system dynamics for the random external force and torque in the predefined range. Figure \ref{fig:slider_crank_full} shows the dynamics response of the dependent coordinates calculated from the minimal coordinates $(\theta_1,\dot \theta_1)$ under the same external force and torque. With the combination of Fig. \ref{fig:slider_crank_full} and Fig. \ref{fig:slider_crank_dyanmics}, we can see that the \updatedTextRebutalTwo{MBD-NODE} provides good-accuracy, long-time prediction for all states. We don't show the coordinates $(y_3,\theta_3,\dot \theta_3)$ because they are zeros for all time.}

\begin{figure}
    \centering
    \includegraphics[width=11cm]{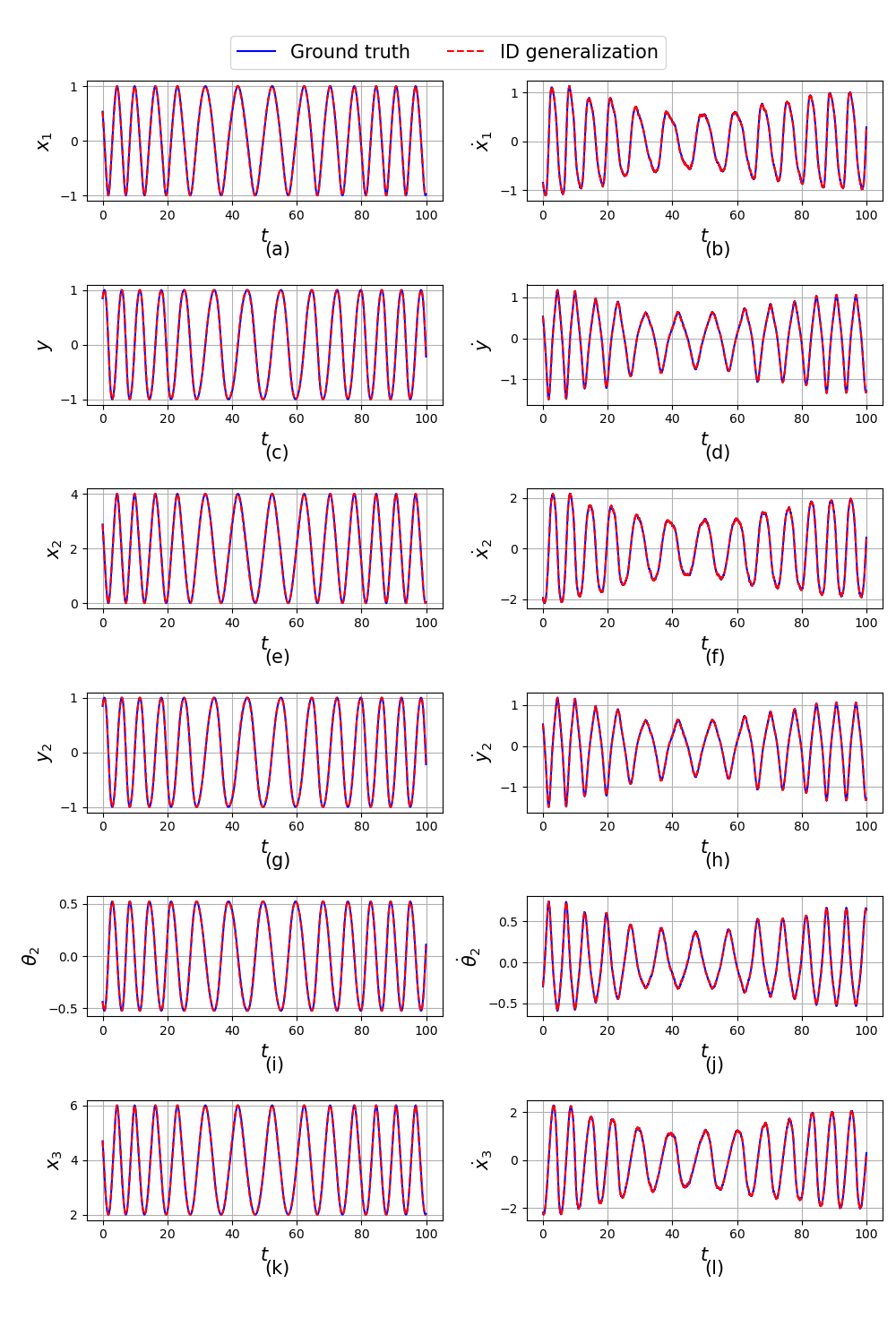}
    \caption{The dynamics response of the slider-crank mechanism under the external force and torque for 10000 time steps. (a): the $x_1$ vs $t$; (b): the $\dot x_1$ vs $t$; (c): the $y_1$ vs $t$; (d): the $\dot y_1$ vs $t$; (e): the $x_2$ vs $t$; (f): the $\dot x_2$ vs $t$; (g): the $y_2$ vs $t$; (h): the $\dot y_2$ vs $t$; (i): the $\theta_2$ vs $t$; (j): the $\dot \theta_2$ vs $t$; (k): the $x_3$ vs $t$; (l): the $\dot x_3$ vs $t$.}
    \label{fig:slider_crank_full}
\end{figure}

\section{Conclusion}
\updatedText{Drawing on the NODE methodology, this work introduces \updatedTextRebutalTwo{MBD-NODE}, a method for the data-driven modeling of MBD problems. 
The performance of \updatedTextRebutalTwo{MBD-NODE} is compared against that of several state-of-the-art data-driven modeling methods by means of seven numerical examples that display attributes encountered in common real-life systems, e.g. energy conservation (single mass-spring system), energy dissipation (single mass-spring-damper system), multiscale dynamics (triple mass-spring-damper system), generalization to different parameters (single pendulum system), MPC-based control problem (cart-pole system), chaotic behavior (double pendulum system), and presence of constraints with long time prediction (slider-crank mechanism). The results demonstrate an overall superior performance of the proposed \updatedTextRebutalTwo{MBD-NODE} method, in the following aspects:}

\begin{enumerate}
    \item \updatedText{Generalization Capability: \updatedTextRebutalTwo{MBD-NODE} demonstrates superior accuracy in both in-distribution (ID) and out-of-distribution (OOD) scenarios, a significant advantage over the ID-focused generalization typically observed with FCNN and LSTM models.}
    \item \updatedText{Model-Based Control Application: The structure of \updatedTextRebutalTwo{MBD-NODE}, mapping system states and external inputs to accelerations, combined with its high generalization accuracy, makes it suitable for model-based control challenges, as demonstrated in the cart-pole control problem.}
    \item \updatedText{Efficiency in Data Usage and Time Independence: Unlike FCNN, \updatedTextRebutalTwo{MBD-NODE}'s integration-based learning does not require extensive time-dependent data, enabling accurate long-term dynamics predictions with less data, as demonstrated in the slider-crank problem.}
    \item \updatedText{Independence from Second-Order Derivative Data: \updatedTextRebutalTwo{MBD-NODE} can predict second-order derivatives based on position and velocity data alone, avoiding the need for direct second-order derivative data required by FCNN and LSTM.}
\end{enumerate}

\updatedText{For reproducibility studies, we provided the open-source code base of \updatedTextRebutalTwo{MBD-NODE}, which includes all the numerical examples and all the trained models used in this study \cite{MNODE_supportData2024}. To the best of our knowledge, this represents the first time mechanical system models and Machine Learning models are made publicly and unrestrictedly  available for reproducibility studies and further research purposes. This can serve as a benchmark testbed for the future development of data-driven modeling methods for multibody dynamics problems.}

\section{\updatedTextRebutalTwo{Limitations and Future Work}}
\updatedTextRebutalTwo{The model proposed has several limitations that remain to be addressed in the future. Firstly, while the extrapolation capabilities of MBD-NODE have been tested on several problems in this work and have shown superior performance compared to traditional models like LSTM and FCNN, additional testing will paint a better picture in relation to the out-of-distribution performance of MBD-NODE. Secondly, although MBD-NODE is efficient in terms of data usage and does not rely on second-order derivative data, as detailed in this contribution, other competing methods come with less computational costs. A study to gauge the MBD-NODE trade-off between computational cost and quality of results would be justified and insightful.}

\updatedTextRebutalTwo{Future work should also focus on optimizing the training process to reduce the MBD-NODE computational costs. Another area for improvement is extending MBD-NODE to work with flexible multibody system dynamics problems. Exploring these directions stands to enhance the practical applicability of MBD-NODE and contribute to its broader adoption.}

\begin{acknowledgements}
This work was carried out in part with support from National Science Foundation project CMMI2153855.
\end{acknowledgements}

%
\section*{Conflict of interest}

The authors declare that they have no known competing financial interests or personal relationships that could have appeared to influence the work reported in this paper.

\bibliographystyle{unsrt}
\bibliography{refsMLMBD}    
\newpage
\appendix
\section{Algorithms for training the MNODE}\label{sec:alg}

\begin{algorithm}[htbp]
\caption{The training algorithm for MBD without constraints.}
\label{alg:training_nocon}
\begin{algorithmic}
\State \textbf{Initialize:} Randomly initialized MNODE $f(\cdot,\mathbf{\Theta})$; choose integrator $\Phi$
\State \textbf{Input:} Ground truth trajectories $\mathcal{T} = \{\mathbf{Z}_{i}\}_{i=0}^{T}$ with parameters $\bm \mu$ and external inputs $\mathbf{u}$, optimizer and its settings
\For{each epoch $e = 1, 2, ..., E$}
    \For{each time step $i = 0, 1, ..., T-1$}
        \State Prepare input state $\mathbf{Z}_i$ and target state $\mathbf{Z}_{i+1}$
        \State Forward pass by integrator $\Phi$ and \updatedText{$f(\cdot,\mathbf{\Theta})$} get the predicted state $\hat{\mathbf{Z}}_{i+1}=\Phi(\mathbf{Z}_i,f,\Delta t_i)$
        \State Compute loss $\text{L} = \|\mathbf{Z}_{i+1} - \mathbf{\hat{Z}}_{i+1}\|_2^2$
        \State Backpropagate the loss to compute gradients $\nabla_{\mathbf{\Theta}}\text{L}$ 
        \State Update the parameters using optimizer: $\mathbf{\Theta} = \text{Optimizer}(\mathbf{\Theta},\nabla_{\mathbf{\Theta}} \text{L})$
    \EndFor
    \State Decay the learning rate using exponential schedule
\EndFor
\State \textbf{Output:} Trained MBD $f(\cdot,\mathbf{\Theta}^*)$
\end{algorithmic}
\end{algorithm}

\begin{algorithm}[htbp]
\caption{The training algorithm for MBD with constraints equation-based optimization and dependent coordinates data.}
\label{alg:training_hard}
\begin{algorithmic}
\State \textbf{Initialize:} Randomly initialized MNODE $f(\cdot,\mathbf{\Theta})$. Choose \updatedText{integrator $\Phi$, identify constraint equation $\phi$ and the map $\phi^{-1}$ from the independent/minimal coordinates to the dependent coordinates.}
\State \textbf{Input:} Ground truth trajectories $\mathcal{T} = \{\mathbf{Z}_{i}\}_{i=0}^{T}$, optimizer and its settings.
\For{each epoch $e = 1, 2, ..., E$}
    \For{each time step $i = 0, 1, ..., T-1$}
        \State \updatedText{Prepare input state $\mathbf{Z}_i=(\mathbf{Z}^{M}_i,\mathbf{Z}^R_i)^T\in \mathbf{R}^{2n_z}$ and target state $\mathbf{Z}_{i+1}=(\mathbf{Z}^{M}_{i+1},\mathbf{Z^R_{i+1}})^T\in \mathbf{R}^{2n_z}$}
        \State \updatedText{Forward pass the minimal coordinates $\mathbf{Z}^{M}_i$ to  $f(\cdot,\mathbf{\Theta})$ with integrator $\Phi$ and to get the predicted minimal coordinates at next time step $\hat{\mathbf{Z}}_{i+1}^{M}=\Phi(\mathbf{Z}_i^{M},f,\Delta t)$}
        \State \updatedText{Recover the dependent coordinates $\hat{\mathbf{Z}}_{i+1}^{R}$ using the independent coordinates  $\hat{\mathbf{Z}}_{i+1}^{M}$ with  $\phi^{-1}$}
        \State \updatedText{Combine the minimal and dependent coordinates to get the full coordinates $\Tilde{\mathbf{Z}}_{i+1}=(\hat{\mathbf{Z}}_{i+1}^{M},\hat{\mathbf{Z}}_{i+1}^{R})^T$}
        \State Compute loss $\text{L} = \|\mathbf{Z}_{i+1} - \mathbf{\Tilde{Z}}_{i+1}\|_2^2$
        \State Backpropagate the loss to compute gradients $\nabla_{\mathbf{\Theta}}\text{L}$ 
        \State Update the parameters using optimizer: $\mathbf{\Theta} = \text{Optimizer}(\mathbf{\Theta},\nabla_{\mathbf{\Theta}} \text{L})$
    \EndFor
    \State Decay the learning rate using exponential schedule
\EndFor
\State \textbf{Output:} Trained MBD $f(\cdot,\mathbf{\Theta}^*)$
\end{algorithmic}
\end{algorithm}

\begin{algorithm}[htbp]
    \caption{The training algorithm for MBD; presence of constraints handled by using minimal/independent coordinates; kinematic constraints used to recover the dependent ones.}
    \label{alg:training_hard_minimal}
    \begin{algorithmic}
    \State \textbf{Initialize:} Randomly initialized MNODE $f(\cdot,\mathbf{\Theta})$\updatedText{; choose integrator $\Phi$; uses pior knowledge of constraint equation $\phi$ and the minimal coordinates.}
    \State \textbf{Input:} Ground truth minimal coordinates trajectories $\mathcal{T} = \{\mathbf{Z}^M_{i}\}_{i=0}^{T}$, optimizer and its settings.
    \For{each epoch $e = 1, 2, ..., E$}
        \For{each time step $i = 0, 1, ..., T-1$}
            \State \updatedText{Prepare input state $\mathbf{Z}_i^M$ and target state $\mathbf{Z}_{i+1}^M$}
            \State \updatedText{Forward pass the minimal coordinates $\mathbf{Z}^{M}_i$ to  $f(\cdot,\mathbf{\Theta})$ with integrator $\Phi$ and to get the predicted minimal coordinates at next time step $\hat{\mathbf{Z}}_{i+1}^{M}=\Phi(\mathbf{Z}_i^{M},f,\Delta t)$}
            \State \updatedText{Compute loss $\text{L} = \|\mathbf{Z}^M_{i+1} - \mathbf{\hat{Z}}^M_{i+1}\|_2^2$}
            \State Backpropagate the loss to compute gradients $\nabla_{\mathbf{\Theta}}\text{L}$ 
            \State Update the parameters using optimizer: $\mathbf{\Theta} = \text{Optimizer}(\mathbf{\Theta},\nabla_{\mathbf{\Theta}} \text{L})$
        \EndFor
        \State Decay the learning rate using exponential schedule
    \EndFor
    \State \textbf{Output:} Trained MBD $f(\cdot,\mathbf{\Theta}^*)$
    \end{algorithmic}
    \end{algorithm}

\clearpage
    
\updatedText{\section{Training time cost for different models and integrators}}\label{appsec:time_cost}
 \begin{table}[htbp]
 \centering
 \begin{tabular}{|m{4cm}|m{2cm}|m{2cm}|m{2.5cm}|}
    \hline
    \textbf{\updatedText{Test Case}} & \textbf{\updatedText{Model }} & \textbf{\updatedText{Integrator}} & \textbf{\updatedText{Time Cost (s)}} \\ \hline
    \multirow{5}{4cm}{\updatedText{Single Mass Spring}} & \updatedText{MNODE} & \updatedText{LF2} & \updatedText{507.73} \\ \cline{2-4} 
    & \updatedText{MNODE} & \updatedText{YS4} & \updatedText{874.05} \\ \cline{2-4} 
    & \updatedText{MNODE} & \updatedText{FK6} & \updatedText{1461.52} \\ \cline{2-4} 
    & \updatedText{HNN} & \updatedText{RK4} & \updatedText{218.02} \\ \cline{2-4} 
    & \updatedText{LNN} & \updatedText{RK4} & \updatedText{988.45} \\ \hline

    \multirow{5}{4cm}{\updatedText{Single Mass Spring Damper}} & \updatedText{MNODE} & \updatedText{FE1} & \updatedText{316.86} \\ \cline{2-4} 
    & \updatedText{MNODE} & \updatedText{MP2} & \updatedText{518.09} \\ \cline{2-4}
    & \updatedText{MNODE} & \updatedText{RK4} & \updatedText{1254.13} \\ \cline{2-4} 
    & \updatedText{FCNN} & \updatedText{-} & \updatedText{220.13} \\ \cline{2-4}
    & \updatedText{LSTM} & \updatedText{-} & \updatedText{500.63} \\ \hline
    \multirow{5}{4cm}{\updatedText{Triple Mass Spring Damper}} &  \updatedText{MNODE} & \updatedText{FE1} & \updatedText{358.55} \\ \cline{2-4}
    &\updatedText{MNODE} & \updatedText{MP2} & \updatedText{608.55} \\ \cline{2-4} 
    & \updatedText{MNODE} & \updatedText{RK4} & \updatedText{915.68} \\ \cline{2-4} 
    & \updatedText{FCNN} & \updatedText{-} & \updatedText{214.13} \\ \cline{2-4}
    & \updatedText{LSTM} & \updatedText{-} & \updatedText{460.33} \\ \hline
   
    \multirow{5}{4cm}{\updatedText{Single Pendulum}} & \updatedText{MNODE} & \updatedText{FE1} & \updatedText{250.52} \\ \cline{2-4} 
    & \updatedText{MNODE} & \updatedText{MP2} & \updatedText{276.12} \\ \cline{2-4} 
    & \updatedText{MNODE} & \updatedText{RK4} & \updatedText{838.37} \\ \cline{2-4} 
    & \updatedText{FCNN} & \updatedText{-} & \updatedText{210.18} \\ \cline{2-4} 
    & \updatedText{LSTM} & \updatedText{-} & \updatedText{348.36} \\ \hline
    
    \multirow{5}{4cm}{\updatedText{Double Pendulum}} 
    & \updatedText{MNODE} & \updatedText{FE1} & \updatedText{253.34} \\ \cline{2-4} 
    & \updatedText{MNODE} & \updatedText{MP2} & \updatedText{368.75} \\ \cline{2-4} 
    & \updatedText{MNODE} & \updatedText{RK4} & \updatedText{854.06} \\ \cline{2-4} 
    & \updatedText{FCNN} & \updatedText{-} & \updatedText{176.51} \\ \cline{2-4} 
    & \updatedText{LSTM} & \updatedText{-} & \updatedText{402.62} \\ \hline
    \multirow{5}{4cm}{\updatedText{Cart-pole}}
    & \updatedText{MNODE} & \updatedText{FE1} & \updatedText{255.80} \\ \cline{2-4} 
    & \updatedText{MNODE} & \updatedText{MP2} & \updatedText{285.04} \\ \cline{2-4} 
    & \updatedText{MNODE} & \updatedText{RK4} & \updatedText{776.08} \\ \cline{2-4} 
    & \updatedText{FCNN} & \updatedText{-} & \updatedText{214.26} \\ \cline{2-4} 
    & \updatedText{LSTM} & \updatedText{-} & \updatedText{358.14} \\ \hline
 \end{tabular}
 \label{tab:time_cost}
 \caption{\updatedText{Time cost for training the models with different integrators. Here the FE1 represents the 1st order Forward Euler, LF2 represents the 2nd order Leapfrog method, MP2 represents the 2nd order Midpoint method, RK4 represents the 4th order Runge-Kutta method, YS4 represents the 4th order Yoshida method, and FK6 represents the 6th order Fukushima method. Please note that compared with the MNODE and LNN, the second-order derivative data is provided to the HNN.}}
\end{table}

\updatedText{\section{The detail formulation of the equation of motion for the slider-crank mechanism}}
\label{appsec:slider_crank}

\updatedText{Following the setting mentioned in Section \ref{sec:slider_crank}, the equation of motion for the slider-crank mechanism can be formulated as follows:}

\updatedText{The mass matrix $M \in \mathbb{R}^{9 \times 9}$ is:}
\updatedText{\begin{equation}
M = \begin{bmatrix}
M_1 & 0_{3\times3} & 0_{3\times3} \\
0_{3\times3} & M_2 & 0_{3\times3} \\
0_{3\times3} & 0_{3\times3} & M_3
\end{bmatrix},
\end{equation}}
\updatedText{where:}
\updatedText{
\begin{equation*}
M_1 = \begin{bmatrix}
m_1 & 0 & 0 \\
0 & m_1 & 0 \\
0 & 0 & I_1
\end{bmatrix} = \begin{bmatrix}
3 & 0 & 0 \\
0 & 3 & 0 \\
0 & 0 & 4
\end{bmatrix},
\end{equation*}
}
\updatedText{
\begin{equation*}
M_2 = \begin{bmatrix}
m_2 & 0 & 0 \\
0 & m_2 & 0 \\
0 & 0 & I_2
\end{bmatrix} = \begin{bmatrix}
6 & 0 & 0 \\
0 & 6 & 0 \\
0 & 0 & 32
\end{bmatrix},
\end{equation*}
}
\updatedText{
\begin{equation*}
M_3 = \begin{bmatrix}
m_3 & 0 & 0 \\
0 & m_3 & 0 \\
0 & 0 & I_3
\end{bmatrix} = \begin{bmatrix}
1 & 0 & 0 \\
0 & 1 & 0 \\
0 & 0 & 1
\end{bmatrix}.
\end{equation*}}
\updatedText{
The states of the slider crank mechanism $(x_1,y_1,\theta_1,x_2,y_2,\theta_2,x_3,y_3,\theta_3)$ follows the below constraints  $\Phi : \mathbb{R}^9 \rightarrow \mathbb{R}^8$ on the position:}
\updatedText{
\begin{equation}
\Phi(q) = \begin{bmatrix}
x_1 - \cos(\theta_1) \\
y_1 - \sin(\theta_1) \\
x_1 + \cos(\theta_1) - x_2 + 2\cos(\theta_2) \\
y_1 + \sin(\theta_1) - y_2 + 2\sin(\theta_2) \\
x_2 + 2\cos(\theta_2) - x_3 \\
y_2 + 2\sin(\theta_2) - y_3 \\
y_3 \\
\theta_3
\end{bmatrix}.
\end{equation}}
\updatedText{
We also have the following constraints $\Phi_q \in \mathbb{R}^{8 \times 9}$on the velocit:}
\updatedText{
\begin{equation}
\Phi_q = \begin{bmatrix}
    1 & 0 & \sin(\theta_1) & 0 & 0 & 0 & 0 & 0 & 0 \\
    0 & 1 & -\cos(\theta_1) & 0 & 0 & 0 & 0 & 0 & 0 \\
    1 & 0 & -\sin(\theta_1) & -1 & 0 & -2\sin(\theta_2) & 0 & 0 & 0 \\
    0 & 1 & \cos(\theta_1) & 0 & -1 & 2\cos(\theta_2) & 0 & 0 & 0 \\
    0 & 0 & 0 & 1 & 0 & -2\sin(\theta_2) & -1 & 0 & 0 \\
    0 & 0 & 0 & 0 & 1 & 2\cos(\theta_2) & 0 & -1 & 0 \\
    0 & 0 & 0 & 0 & 0 & 0 & 0 & 1 & 0 \\
    0 & 0 & 0 & 0 & 0 & 0 & 0 & 0 & 1 \\
    \end{bmatrix}.
    \end{equation}
}
\updatedText{
The vector $F_e \in \mathbb{R}^{9}$ from external forces is:
\begin{equation}
F_e = \begin{bmatrix}
    F_{e1}\\
    F_{e2}\\
    F_{e3}
    \end{bmatrix},
\end{equation}
}
\updatedText{where:}
\updatedText{ 
    \begin{equation*}
    F_{e1} = \begin{bmatrix}
        0 \\
        0 \\
        T 
        \end{bmatrix} \in \mathbb{R}^{3},
        \end{equation*}}
\updatedText{ 
                \begin{equation*}
                    F_{e2} = \begin{bmatrix}
                        0 \\
                        0 \\
                        0 
                        \end{bmatrix} \in \mathbb{R}^{3},
                        \end{equation*}}
\updatedText{
                        \begin{equation*}
                            F_{e3} = \begin{bmatrix}
                                F \\
                                0 \\
                                0
                                \end{bmatrix} \in \mathbb{R}^{3}.
                                \end{equation*}}

\updatedText{We can rearrange the constraint equations on the acceleration:}
\updatedText{
    \begin{subequations}
    \begin{align}
        \ddot{x}_1 + \ddot{\theta}_1 \sin(\theta_1) + \dot{\theta}_1^2 \cos(\theta_1) &= 0 \\
        \ddot{y}_1 - \ddot{\theta}_1 \cos(\theta_1) + \dot{\theta}_1^2 \sin(\theta_1) &= 0 \\
        \ddot{x}_1 - \ddot{\theta}_1 \sin(\theta_1) - \dot{\theta}_1^2 \cos(\theta_1) - \ddot{x}_2 - 2 \ddot{\theta}_2 \sin(\theta_2) - 2 \dot{\theta}_2^2 \cos(\theta_2) &= 0 \\
        \ddot{y}_1 + \ddot{\theta}_1 \cos(\theta_1) - \dot{\theta}_1^2 \sin(\theta_1) - \ddot{y}_2 + 2 \ddot{\theta}_2 \cos(\theta_2) - 2 \dot{\theta}_2^2 \sin(\theta_2) &= 0 \\
        \ddot{x}_2 - 2 \ddot{\theta}_2 \sin(\theta_2) - 2 \dot{\theta}_2^2 \cos(\theta_2) - \ddot{x}_3 &= 0 \\
        \ddot{y}_2 + 2 \ddot{\theta}_2 \cos(\theta_2) - 2 \dot{\theta}_2^2 \sin(\theta_2) - \ddot{y}_3 &= 0 \\
        \dot{\theta}_3 &= 0 \\
        \ddot{\theta}_3 &= 0
        \end{align}
        \end{subequations}
}
\updatedText{ 
to get the $\gamma_c$ as:}
\updatedText{ 
        \begin{equation}
            \gamma_c =\begin{bmatrix}
                -\dot{\theta}_1^2 \cos(\theta_1) \\
                -\dot{\theta}_1^2 \sin(\theta_1) \\
                \dot{\theta}_1^2 \cos(\theta_1) + 2\dot{\theta}_2^2 \cos(\theta_2) \\
                \dot{\theta}_1^2 \sin(\theta_1) + 2\dot{\theta}_2^2 \sin(\theta_2) \\
                2\dot{\theta}_2^2 \cos(\theta_2) \\
                2\dot{\theta}_2^2 \sin(\theta_2) \\
                0 \\
                0
                \end{bmatrix}.
                \end{equation}}
\updatedText{Finally, we plug the above equations into Eq. \ref{eqn:MBD} to get the equation of motion for the slider-crank mechanism.}
\end{document}